\documentclass[twocolumn,tighten]{aastex63}
\usepackage{amsmath}
\usepackage{natbib}
\usepackage{diagbox}

\shorttitle{TDE simulations with accurate stellar structures}
\shortauthors{Golightly et al.}

\begin{document}

\title{On the Diversity of Fallback Rates from Tidal Disruption Events with Accurate Stellar Structure}

\correspondingauthor{C.~J.~Nixon}
\email{cjn@leicester.ac.uk}

\author{E.~C.~A.~Golightly}
\affiliation{Department of Physics and Astronomy, University of Leicester, Leicester, LE1 7RH, UK}

\author[0000-0002-2137-4146]{C.~J.~Nixon}
\affiliation{Department of Physics and Astronomy, University of Leicester, Leicester, LE1 7RH, UK}

\author[0000-0003-3765-6401]{E.~R.~Coughlin}
\affiliation{Department of Astrophysical Sciences, Peyton Hall, Princeton University, Princeton, NJ 08544}
\affiliation{Columbia Astrophysics Laboratory, New York, NY 80980}

\begin{abstract}
The tidal disruption of stars by supermassive black holes (SMBHs) can be used to probe the SMBH mass function, the properties of individual stars, and stellar dynamics in galactic nuclei. Upcoming missions will detect thousands of TDEs, and accurate theoretical modeling is required to interpret the data with precision. Here we analyze the influence of more realistic stellar structure on the outcome of TDEs; in particular, we compare the fallback rates -- being the rate at which tidally-disrupted debris returns to the black hole -- from progenitors generated with the stellar evolution code {\sc mesa} to ${\gamma \simeq 4/3}$ and $\gamma = 5/3$ polytropes. We find that {\sc mesa}-generated density profiles yield qualitatively-different fallback rates as compared to polytropic approximations, and that only the fallback curves from low-mass ($1M_{\odot}$ or less), zero-age main-sequence stars are well fit by either a ${\gamma \simeq 4/3}$ or $5/3$ polytrope. Stellar age has a strong affect on the shape of the fallback curve, and can produce characteristic timescales (e.g., the time to the peak of the fallback rate) that greatly differ from the polytropic values. We use these differences to assess the degree to which the inferred black hole mass from the observed lightcurve can deviate from the true value, and find that the discrepancy can be at the order of magnitude level. Accurate stellar structure also leads to a substantial variation in the critical impact parameter at which the star is fully disrupted, and can increase the susceptibility of the debris stream to fragmentation under its own self-gravity. These results suggest that detailed modeling is required to accurately interpret observed lightcurves of TDEs.
\end{abstract}


\section{Introduction}
The tidal destruction of a star by a supermassive black hole (SMBH) generates a stream of stellar debris that, over timescales of months to years, feeds the SMBH and generates a luminous, observable signature \citep{hills75, lacy82, Rees:1988aa}. These tidal disruption events (TDEs) therefore offer one of the few means to directly probe the inner regions of otherwise-quiescent galaxies, and dozens have now been observed (e.g., \citealt{gezari12, chornock14, arcavi14, blagorodnova17, hung17, vanvelzen19}; see \citealt{Komossa:2015aa} for a review). However, our ability to confidently use the observed flares from TDEs to study, for example, black hole demographics depends critically on our physical understanding of the disruption process, and in particular the way in which the properties of the progenitor star translate to a corresponding feeding rate of the SMBH.

Along these lines, \citet{Lodato:2009aa} simulated the full disruption of polytropes, which offer simple, yet physical descriptions of the density profiles of stellar interiors (e.g., \citealt{hansen04}), and also provided a nearly-analytical means of calculating the fallback rate from a star with a given density profile (using the impulse, or ``frozen-in'', approximation; see also \citealt{stone13} and Section \ref{sec:impulse} below). They found that, while the fallback rate always approached the expected, $t^{-5/3}$ scaling at late times, the peak value of the fallback and the time to peak depended on the stellar structure. \citet{Guillochon:2013aa} expanded this work by also studying the effect of varying the point of closest approach of the stellar center of mass to the black hole, and found that denser stars more frequently leave bound ``cores'' that either resist the tidal shear altogether throughout pericenter passage, or reform following the full disruption of the star. The existence of these cores then modifies not only the early-time fallback, but also causes the late-time fallback to deviate from $t^{-5/3}$, and these features are a direct result of the stellar structure (in combination with the variation in the stellar pericenter).

Since then, a number of other authors have analyzed the tidal disruption of polytropes, with the aim of assessing one or another aspect of the tidal disruption process (e.g. \citealt{hayasaki13,Coughlin:2015aa,Shiokawa:2015aa,Bonnerot:2016aa,Hayasaki:2016aa,Coughlin:2016aa,mainetti17,guillochon17,bonnerot17,Coughlin:2017aa,Wu:2018aa,Golightly:2019aa}). However, only relatively few authors have analyzed the disruption of a star with a density profile \emph{other than} a polytrope. Of which we are aware, those studies are \citet{macleod12}, who investigated the disruption of giant stars by particularly massive SMBHs; \citet{law-smith17}, who simulated the disruption of white dwarfs with extended, hydrogen envelopes; \citet{Gallegos-Garcia:2018aa}, who analytically calculated the fallback of metal-rich material from the cores of evolved stars; and \citet{Goicovic:2019aa}, who analyzed the extent to which a more realistic stellar structure affects the stellar ``disruptability''.

Among the questions that remain concerning stellar structure is the degree to which more realistic stellar density profiles (i.e., those calculated with a stellar evolution code that accounts for more realistic opacities, metallicity gradients, and energy transport) affects the fallback rate \emph{compared to polytropes}. More realistic stellar profiles could impose additional variability on the fallback rate, or conceivably alter characteristic timescales associated with the fallback (e.g., the time to the peak of the fallback curve). Such timescales are used to place constraints on black hole properties (e.g., \citealt{Guillochon:2013aa, mockler19}), and hence it is necessary to understand the effects that stellar structure can have in modifying them.

In this paper we analyze the disruption of stars with stellar structure computed with the code {\sc mesa} \citep{Paxton:2011aa,Paxton:2013aa,Paxton:2015aa,Paxton:2018aa}, primarily to understand the influence that such structure can have on the fallback of debris to the SMBH. In Section \ref{sec:stellar}, we first describe and present the stellar profiles calculated with {\sc mesa}, and in Section \ref{sec:impulse} we use the impulse approximation -- as described in \citet{Lodato:2009aa} -- to calculate the fallback rate onto the black hole from those stars; we show that, compared to polytropes that are matched to the same stellar mass and radius of a given {\sc mesa} model, there are certain combinations of initial stellar mass and age that yield notably different fallback curves. In Section \ref{sec:numerical} we present the results of numerical simulations of disruptions of the {\sc mesa} models, and compare those results to disruptions of polytropes (again, matched to the same stellar mass and radius). We discuss the implications of our simulations for estimating the black hole mass from observed lightcurves in Section~\ref{sec:massestimate}, and show that polytropic approximations can lead to mass-estimate discrepancies at the order of magnitude level. We summarize and conclude in Section \ref{sec:conclusions}.

\section{Stellar profiles}
\label{sec:stellar}
Using the stellar evolution code {\sc mesa} \citep{Paxton:2011aa,Paxton:2013aa,Paxton:2015aa,Paxton:2018aa}, we evolved a 0.3 $M_{\odot}$, 1$M_{\odot}$, and 3$M_{\odot}$ zero-age main sequence (ZAMS) star to the end of the main sequence, being the time at which the hydrogen mass fraction in the core dropped below 0.1\%. For each star, we used all of the default values for the standard inputs (e.g., each pre-main sequence model adopted Solar metallicity, the stars were all non-rotating, there was no mass loss in the form of winds) within {\sc mesa}, version 10398.

\begin{figure*}
   \centering
   \includegraphics[width=0.32\textwidth]{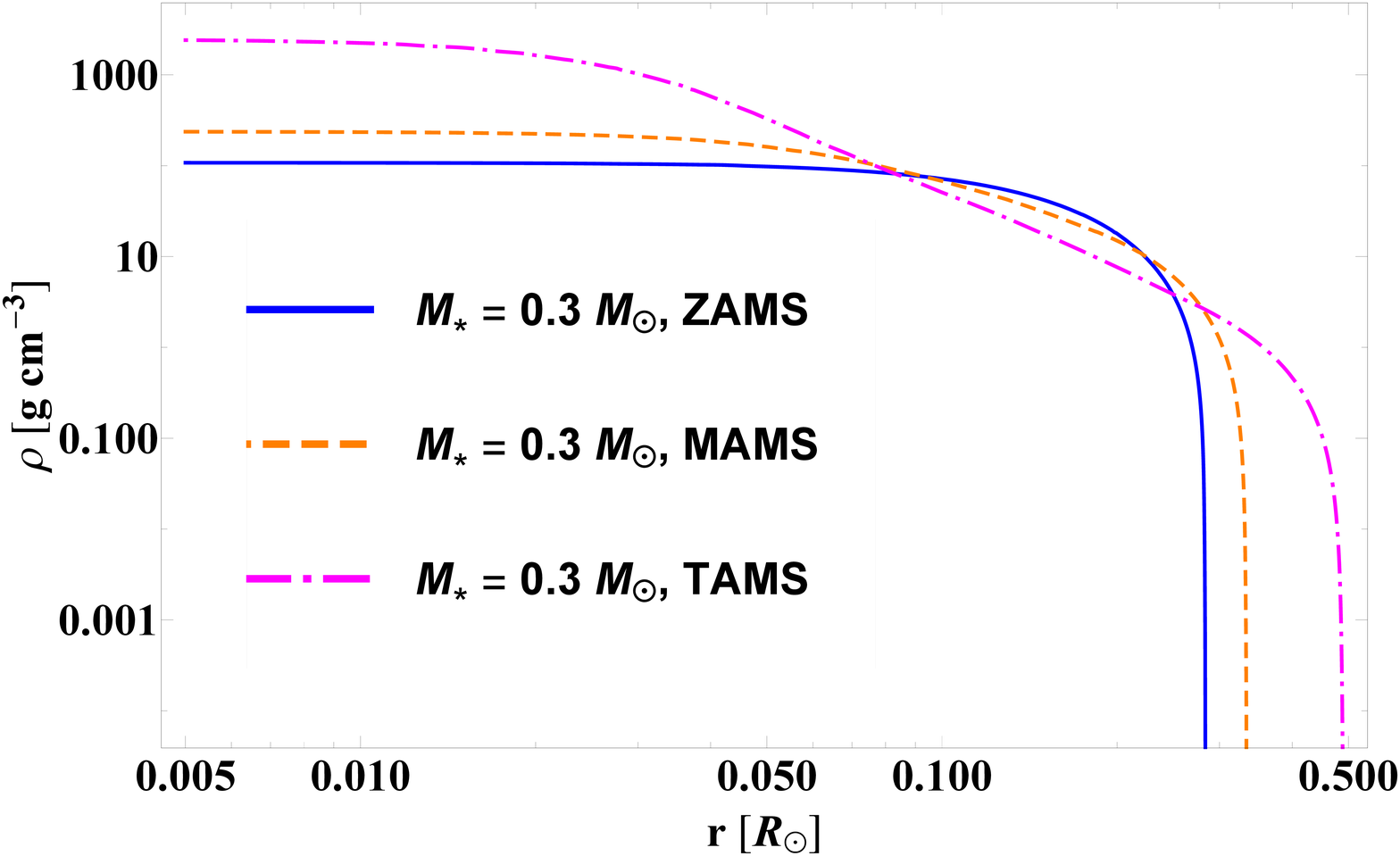}
   \includegraphics[width=0.32\textwidth]{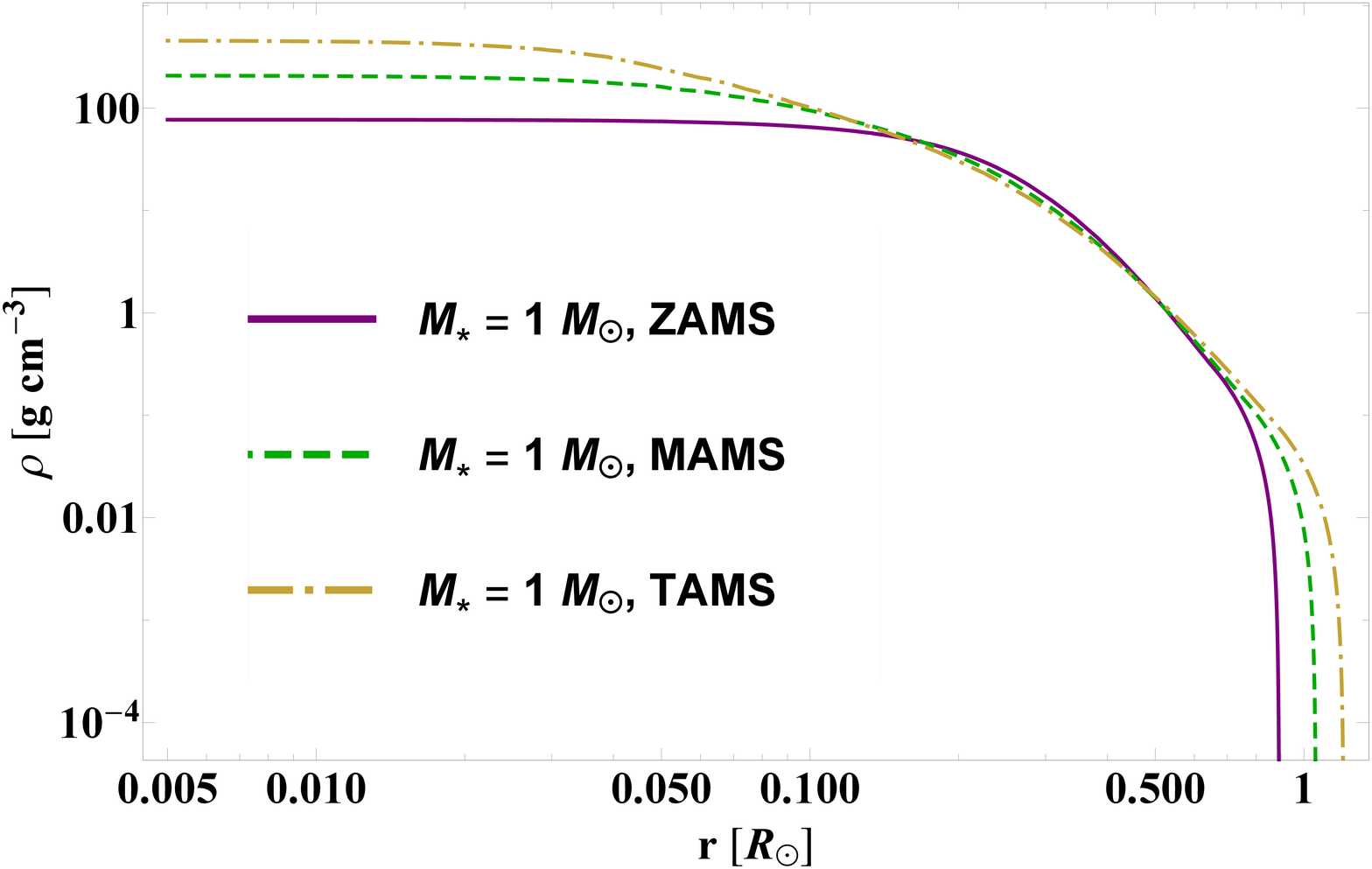}
   \includegraphics[width=0.32\textwidth]{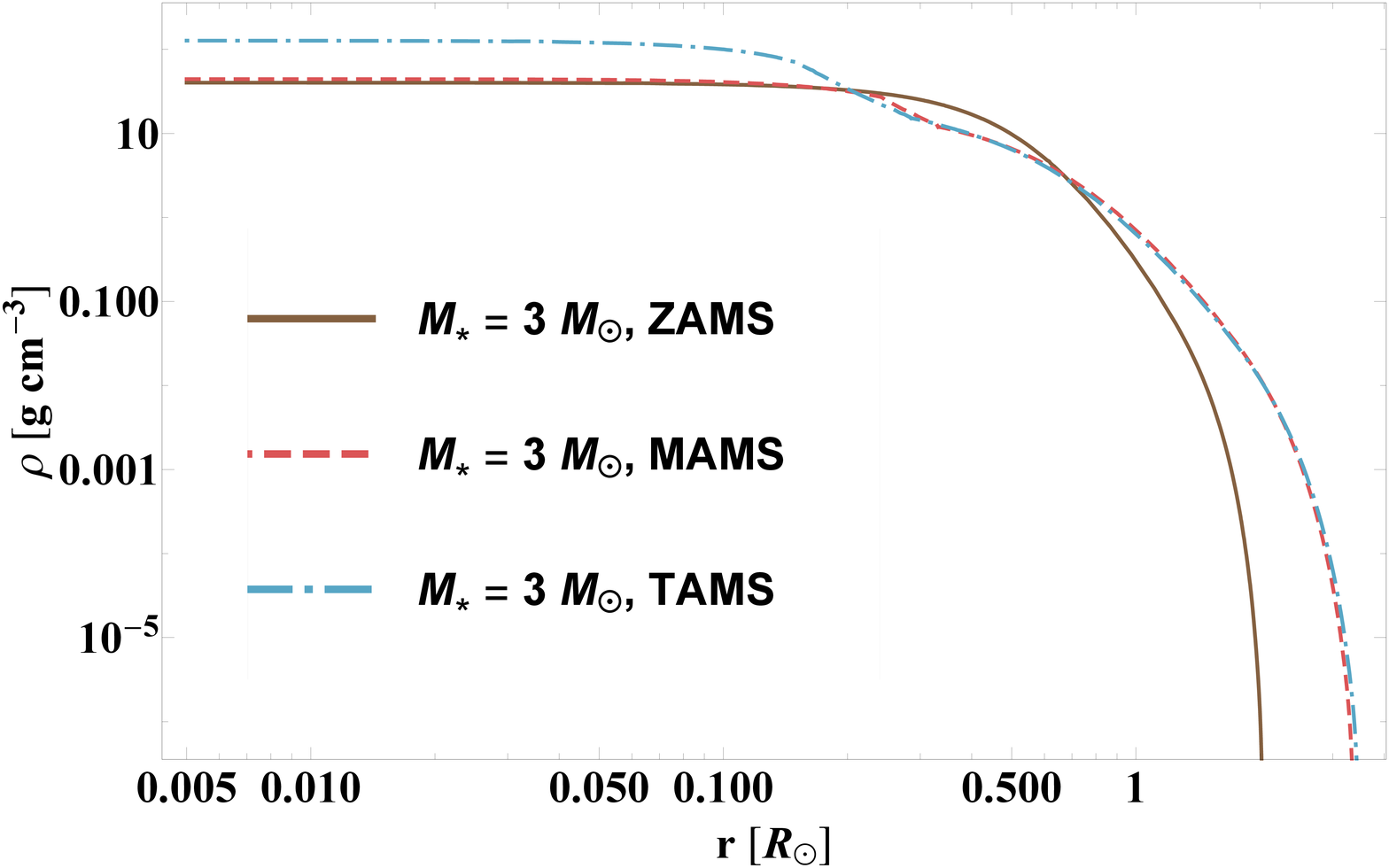} 
   \caption{Stellar density profiles computed with {\sc mesa} in g cm$^{-3}$ as a function of radius in Solar radii. The left panel shows the 0.3 $M_{\odot}$ star, the middle the 1$M_{\odot}$ star, and the right the 3 $M_{\odot}$ star, with the different ages shown by the different lines as indicated in the legends.}
   \label{fig:dens}
\end{figure*}

We took snapshots of the density of each star at the zero-age main sequence (ZAMS), the terminal-age main sequence (TAMS), and at one time in between ZAMS and TAMS when the hydrogen mass fraction in the core just fell below 0.2; we denote the latter by a ``middle-age main sequence'' star, or MAMS\footnote{We are aware that 0.3\,$M_\odot$ stars have not reached TAMS within a Hubble time. Nonetheless, the density profile of such a star could be achieved at an earlier epoch by a more massive progenitor with, e.g., vigorous mass loss in the form of winds or a more metal-rich environment.}. Figure~\ref{fig:dens} shows the density profile of each of these stars, and demonstrates, perhaps not surprisingly, that stellar evolution produces vastly different density structures over the lifetime of a given star.

In the following two sections, we describe two different approaches to modeling the tidal disruption of these stars by an SMBH. 

\section{The impulse approximation}
\label{sec:impulse}
A useful methodology for analyzing the fallback of debris from a tidal disruption event is the impulse approximation, which posits that the tidal field of the black hole acts impulsively as the center of mass of the star reaches the tidal radius. Therefore, prior to reaching the tidal radius the star retains perfect hydrostatic balance, and thereafter the star is ``destroyed'', meaning that each gas parcel follows its own ballistic orbit in the gravitational field of the black hole. Within this approximation and to lowest order in the tidal potential, the binding energy of a given fluid parcel is only a function of the projected distance of that fluid parcel from the black hole onto the line connecting the stellar center of mass and the black hole. This energy dependence implies that perpendicular ``slices'' of the star return simultaneously to the black hole, which allows one to construct a fallback rate $\dot{M}$ -- being the rate at which bound material returns to the black hole -- that accounts for the stellar density profile $\rho$; the result is \citep{Lodato:2009aa,Gallegos-Garcia:2018aa,Golightly:2019aa}

\begin{equation}
\dot{M} = \frac{M_\star}{T_{\rm mb}}\tau^{-5/3}\int_{\tau^{-2/3}}^{1}\frac{\rho (x) x dx}{\rho_\star}, \label{mdot}
\end{equation}
where $M_\star$ is the stellar mass, $\tau = t/T_{\rm mb}$ with

\begin{equation}
T_{\rm mb} = \left(\frac{R_\star}{2}\right)^{3/2}\frac{2\pi M}{M_\star\sqrt{GM}}
\end{equation}
the return time of the most bound debris, $M$ the black hole mass and $R_\star$ the stellar radius, $\eta = R/R_\star$ with $R$ the spherical distance from the center of the star, and $\rho_\star = 3M_\star/(4\pi R_\star^3)$ is the average stellar density. Note that the integral in Equation \eqref{mdot} is negative for all times $t < T_{\rm mb}$, and hence the physical fallback rate is zero until the most bound debris element returns to the black hole.

From Equation \eqref{mdot}, within the impulse approximation the only timescale relevant to the problem is the return time of the most bound debris; hence the time at which the fallback reaches any characteristic value (e.g., the time to peak, the time between half-peaks, the time to reach $\dot{M} \propto t^{-5/3+\alpha}$ with $\alpha > 0$) is also a constant multiple -- which depends on the stellar structure -- of this timescale. If we denote the dimensionless time to any characteristic fallback value by $\tau_{\rm c}$, then by definition the corresponding physical time $t_{\rm c}$ is

\begin{equation}
t_{\rm c} = \left(\frac{R_\star}{2}\right)^{3/2}\frac{2\pi M}{M_\star\sqrt{GM}} \tau_{\rm c}(\rho).
\end{equation}
This simple expression demonstrates the relative importance that stellar structure can have when \emph{inferring} black hole properties from observations\footnote{We assume for simplicity that the observed accretion luminosity scales with the fallback rate; one can permit further flexibility in this regard, but only at the expense of introducing additional uncertainties.}: assume that for a given tidal disruption event with known $t_{\rm c}$, we know the mass and radius of the disrupted star. Then for the same event if we ascribe to the disrupted star two different density profiles, $\rho_1$ and $\rho_{2}$, then we will infer two different black hole masses for the same event, $M_1$ and $M_2$, their ratio being

\begin{equation}
\frac{M_2}{M_1} = \left(\frac{\tau_{\rm c}(\rho_1)}{\tau_{\rm c}(\rho_2)}\right)^2. \label{M2M1}
\end{equation}
Thus, relatively small differences in the stellar structure are capable of producing more discrepant black hole masses owing to the squared dependence in this expression.

\begin{figure}
   \centering
   \includegraphics[width=\columnwidth]{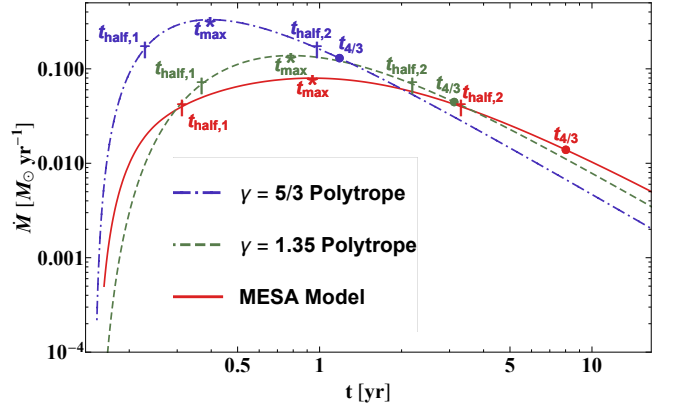} 
   \caption{The fallback rate calculated using the frozen-in approximation, in Solar masses per year as a function of time in years, for the 1$M_{\odot}$, TAMS star, when the density profile is modeled as a $\gamma = 5/3$ polytrope (blue, dot-dashed) and a $\gamma = 1.35$ polytrope (green, dashed); the red, solid curve shows the fallback rate calculated from the {\sc mesa} progenitor. Here the black hole mass was set to $10^{6}M_{\odot}$. The different points show characteristic times in the fallback rate, including the time taken to reach the peak ($t_{\rm max}$, asterisks), to reach $\dot{M}\propto t^{-4/3}$ ($t_{4/3}$, bullets), and the time taken to reach half the peak fallback rate (which occurs on both the rise and the decay of the curve; respectively $t_{\rm half,1}$ and $t_{\rm half,2}$, crosses). It is apparent that, while each one of these stars possesses the same mass and radius, the density profile also plays a large role in generating differences between these characteristic timescales.}
   \label{fig:mdots1}
\end{figure}

Figure \ref{fig:mdots1} illustrates the fallback rate onto the black hole (in units of Solar masses per year as a function of time in years) from the frozen-in approximation, where the disrupted star has a mass and radius equal to those of the $1 M_{\odot}$, TAMS star (see Table \ref{tab:mesa_properties}) and the black hole has a mass of $10^{6}M_{\odot}$. Each curve adopts a different stellar density profile, being a $\gamma = 5/3$ polytrope (blue, dot-dashed), a $\gamma = 1.35$ polytrope (green, dashed), and the profile resulting from {\sc mesa} (red, solid). It is clear from this figure that, despite the fact that the bulk stellar properties are the same, the density profile has a marked effect on the shape of the curve. To highlight the differences induced by stellar structure, the points on each curve give the time to different, characteristic values of the fallback curve, being the time to peak ($t_{\rm max}$), the time to reach half the peak (which occurs on the rise and the decay of the curve, denoted respectively by $t_{\rm half,1}$ and $t_{\rm half,2}$), and the time to reach $\dot{M} \propto t^{-4/3}$, $t_{4/3}$. While certain characteristic times are visibly comparable for each model (e.g., $t_{\rm max}$), other timescales are more noticeably discrepant (e.g., $t_{4/3}$). 

Because of these discrepancies, for a TDE with a given, physical timescale, the black hole mass required to yield a fallback curve with that timescale will differ according to the stellar structure model that one employs, and the magnitude of the difference in mass will be larger or smaller depending on the characteristic timescale itself. For example, using Equation \eqref{M2M1} and letting $\rho_1$ describe the density profile of the $1M_{\odot}$, TAMS {\sc mesa} progenitor, we find $M_2/M_1 \simeq 2.3$ if one models the density profile by a $\gamma = 1.35$ polytrope and uses the difference between the time to peak and the first time to half peak (i.e., the timescale $t_{\rm c} = t_{\rm max} - t_{\rm half,1}$\footnote{For an observed TDE, one does not know the time at which the star was disrupted, and hence the reference time should also be set to some physical timescale associated with the fallback.}); physically, one can interpret this result by saying that a $\gamma = 1.35$ polytrope requires a black hole mass $\sim 2.3$ times larger to reproduce the time to peak from the $1M_{\odot}$, TAMS {\sc mesa} progenitor. On the other hand, if we use the same timescale but model the star as a $\gamma = 5/3$ polytrope, then we find $M_2/M_1 \simeq 14$ -- because a $\gamma = 5/3$ polytrope peaks considerably earlier than the real star, a much larger black hole mass is required in order to extend the time to maximum fallback. 

\begin{figure}
   \centering
   \includegraphics[width=\columnwidth]{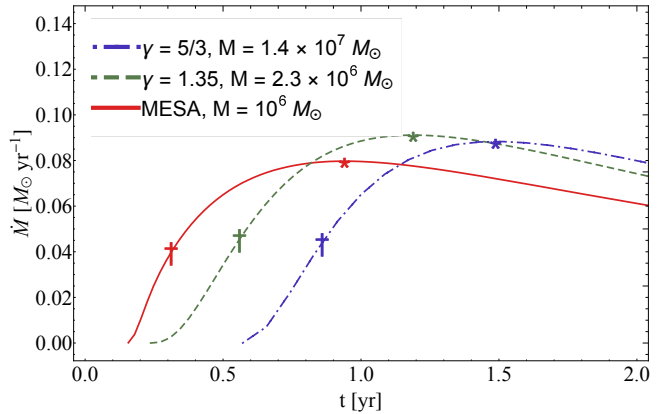} 
   \caption{The fallback rate in Solar masses per year, as a function of time in years on a linear-linear scale, for the $1M_{\odot}$, TAMS {\sc mesa} progenitor, shown by the red, solid curve, when the star is disrupted by a SMBH with mass $M_{} = 10^{6}M_{\odot}$ (and is therefore the same as the red, solid curve in Figure \ref{fig:mdots1}). The green, dashed curve and the blue, dot-dashed curve show the fallback rates for a $\gamma = 1.35$ and $\gamma = 5/3$ polytrope, respectively, when the stellar properties (mass and radius) are matched to those of the {\sc mesa} progenitor. By changing the black hole mass to the value shown in the legend, we are able to reproduce the time taken to go from half peak to peak, being $t_{\rm c} = 0.63$ yr. {}{For each curve, this timescale is the time taken to go from the $\dagger$ to the $\star$.} }
   \label{fig:mdots_tc}
\end{figure}

Figure \ref{fig:mdots_tc} directly demonstrates the way by which changing the mass of the black hole yields commensurate characteristic timescales: the solid, red curve shows the same fallback rate as in Figure \ref{fig:mdots1} for the $1M_{\odot}$, TAMS stellar profile computed with {\sc mesa} and a $10^{6}M_{\odot}$ SMBH (note that this figure is on a linear-linear scale). The dashed, green line and the dot-dashed, blue line show the fallback rates for a $\gamma = 1.35$ and $\gamma = 5/3$ polytrope, respectively, again with the same stellar properties as those of the {\sc mesa} progenitor. However, here we varied the mass of the black hole according to the value required to yield the same time to peak, being $M_{} = 2.3\times 10^{6}$ for the $\gamma = 1.35$ polytrope and $M_{} = 1.4\times 10^{7} M_{\odot}$ for the $\gamma = 5/3$ polytrope. {{} The time between the first half-peak (marked with a $\dagger$) and the peak fallback rate (marked with a $\star$) is the same in each case, showing} that -- by varying the black hole mass appropriately -- one can reconstruct the same, physical timescale for different stellar properties.

One can repeat this procedure for the nine different models presented in Table \ref{tab:mesa_properties} and thereby assess the degree to which a polytropic density profile reproduces the fallback curve -- and correspondingly the inferred black hole mass -- of the one obtained with the {\sc mesa} model. However, the impulse approximation ignores crucial physics of the disruption process that also alter the fallback rate \citep{Guillochon:2013aa, Coughlin:2015aa, steinberg19}. In the next section, we employ hydrodynamical simulations to obtain more realistic fallback rates.

\section{Hydrodynamic simulations}
\label{sec:numerical}
\subsection{Setup}
We use the Smoothed Particle Hydrodynamics (SPH) code {\sc phantom} \citep{Price:2018aa} to model the hydrodynamics of the disruption process. Following previous work \citep[e.g.,][]{Coughlin:2015aa,Coughlin:2016aa,Wu:2018aa,Golightly:2019aa} we model the central SMBH as a Newtonian point mass at the origin with an accretion radius, inside which particles are removed from the simulation. We include the self-gravity of the star, and we model the stellar pressure using an adiabatic equation of state $P=K\rho^\gamma$ where $K$ is a conserved quantity for each fluid element, but can be spatially dependent within the original star. $K$ is chosen such that the fluid pressure is the total pressure in the {\sc mesa} calculation, and thus the SPH star has an equilibrium density structure that matches the stellar density structure calculated by {\sc mesa}. The properties of each star are given in Table~\ref{tab:mesa_properties}. 

We construct the star with particles placed on a close-packed sphere that is stretched to achieve the desired density distribution. We then relax the particle distribution in isolation with a velocity damping force until the star settles into a numerically-relaxed configuration. We plot the density structure at this point against the solutions from {\sc mesa} in the Appendix (Fig.~\ref{fig:density}). We further checked that these solutions are numerically-relaxed by evolving them in isolation for the time taken for the star to reach pericentre in each case, and we found no subsequent evolution of the density profile. 

\begin{table}
\begin{center}
\begin{tabular}{|c|c|c|c|c|} \hline
Star & Mass ($M_{\odot}$) & Radius ($R_{\odot}$) & $r_{\rm t} (R_{\rm g})$\\
\hline
$0.3 M_{\odot}$ ZAMS & $0.3 $ & 0.28 & 20\\
\hline
$0.3 M_{\odot}$ MAMS & $0.3$ & 0.33 & 24\\
\hline
$0.3 M_{\odot}$ TAMS & $0.3$ & 0.49 & 34\\
\hline
$1 M_{\odot}$ ZAMS & $1.0$ & 0.89 & 42\\
\hline
$1 M_{\odot}$ MAMS & $1.0$ & 1.1 & 50\\
\hline
$1 M_{\odot}$ TAMS & $1.0$ & 1.2  & 57\\
\hline
$3 M_{\odot}$ ZAMS &$3.0$ & 2.0 & 67\\
\hline
$3 M_{\odot}$ MAMS & $3.0$ & 3.4 & 112\\
\hline
$3 M_{\odot}$ TAMS & $3.0$ & 3.5 & 115\\
\hline
\end{tabular}
\caption{The properties of each star evolved in {\sc mesa}, being, from left to right, the star name, stellar mass $M_{\star}$ in Solar masses, stellar radius $R_{\star}$ in Solar radii, and tidal radius ($r_{\rm t} \equiv R_{\star}\left(M/M_{\star}\right)^{1/3}$), where $M = 10^{6}M_{\odot}$, in units of gravitational radii of a $10^{6}M_{\odot}$ SMBH. Each disruption has $\beta = 3$, and hence the pericenter distance of each star to the SMBH is obtained by dividing the tidal radius (fourth column) by 3.}
\label{tab:mesa_properties}
\end{center}
\end{table}

We then placed the relaxed stars on parabolic orbits around a central SMBH, with the initial location at five tidal radii. To ensure full disruption of at least some of the stars we chose an impact parameter of $\beta\equiv r_{\rm t}/r_{\rm p} = 3$, where $r_{\rm t} = R_{\star}(M/M_\star)^{1/3}$ is calculated from the mass and radius of each star (see Table \ref{tab:mesa_properties}) and $r_{\rm p}$ is the pericenter distance of the stellar center of mass. Thus, while the $\beta$ is the same for each simulation (and therefore the average strength of the tidal field at pericenter is the same), the physical pericenter is different from simulation to simulation. As we shall see below, $\beta = 3$ is still not large enough to completely disrupt all of the {\sc mesa}-generated stars, but we neglected to go to higher $\beta$ because of the prohibitively-large particle number required to achieve converged results. For each simulation we employed $10^6$ particles, though we also ran tests with $10^5$ particles and found only small differences (at the noise level) in the fallback. 

\begin{figure*}[h!]
   \centering
   \includegraphics[width=0.495\textwidth]{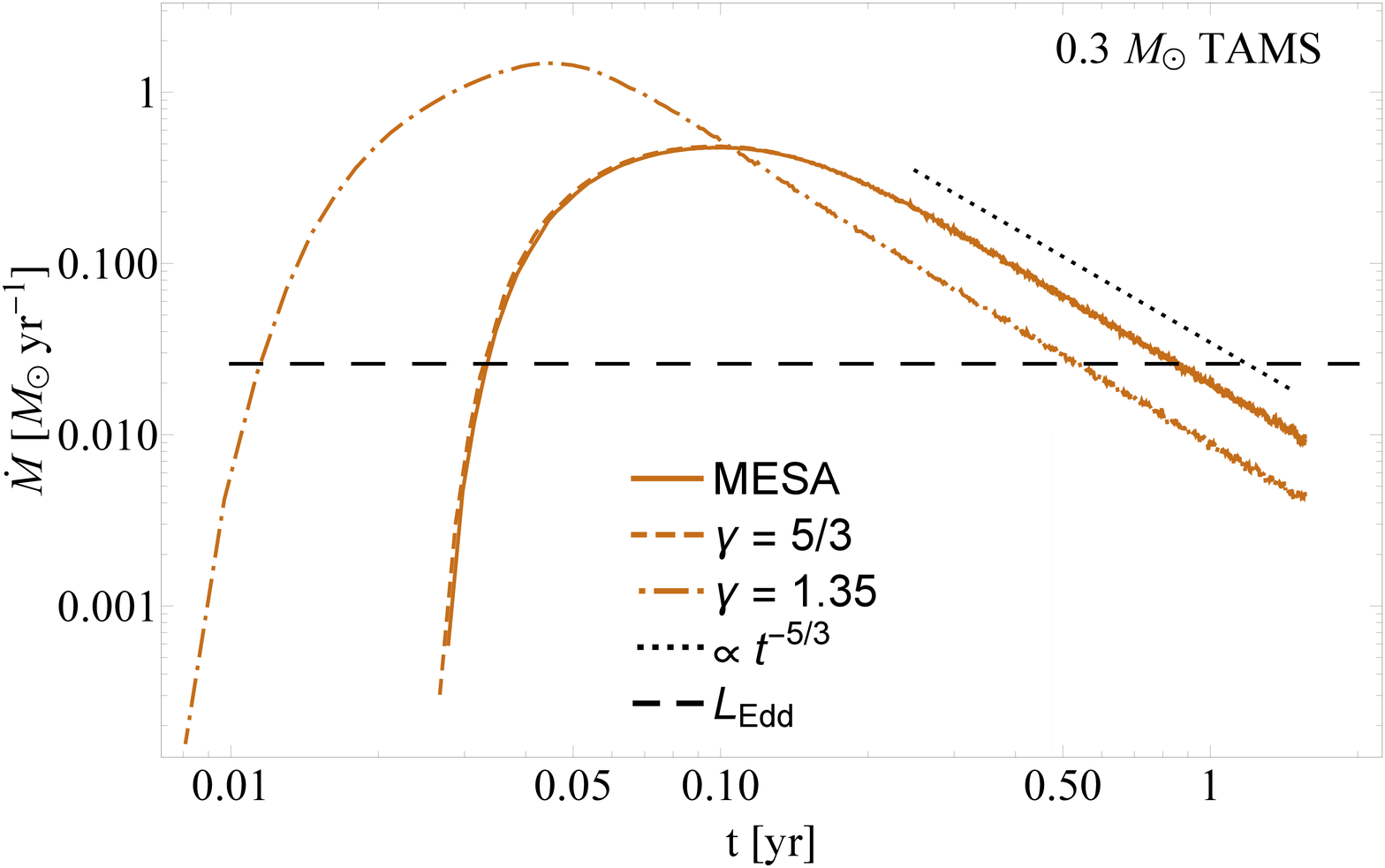}
   \includegraphics[width=0.495\textwidth]{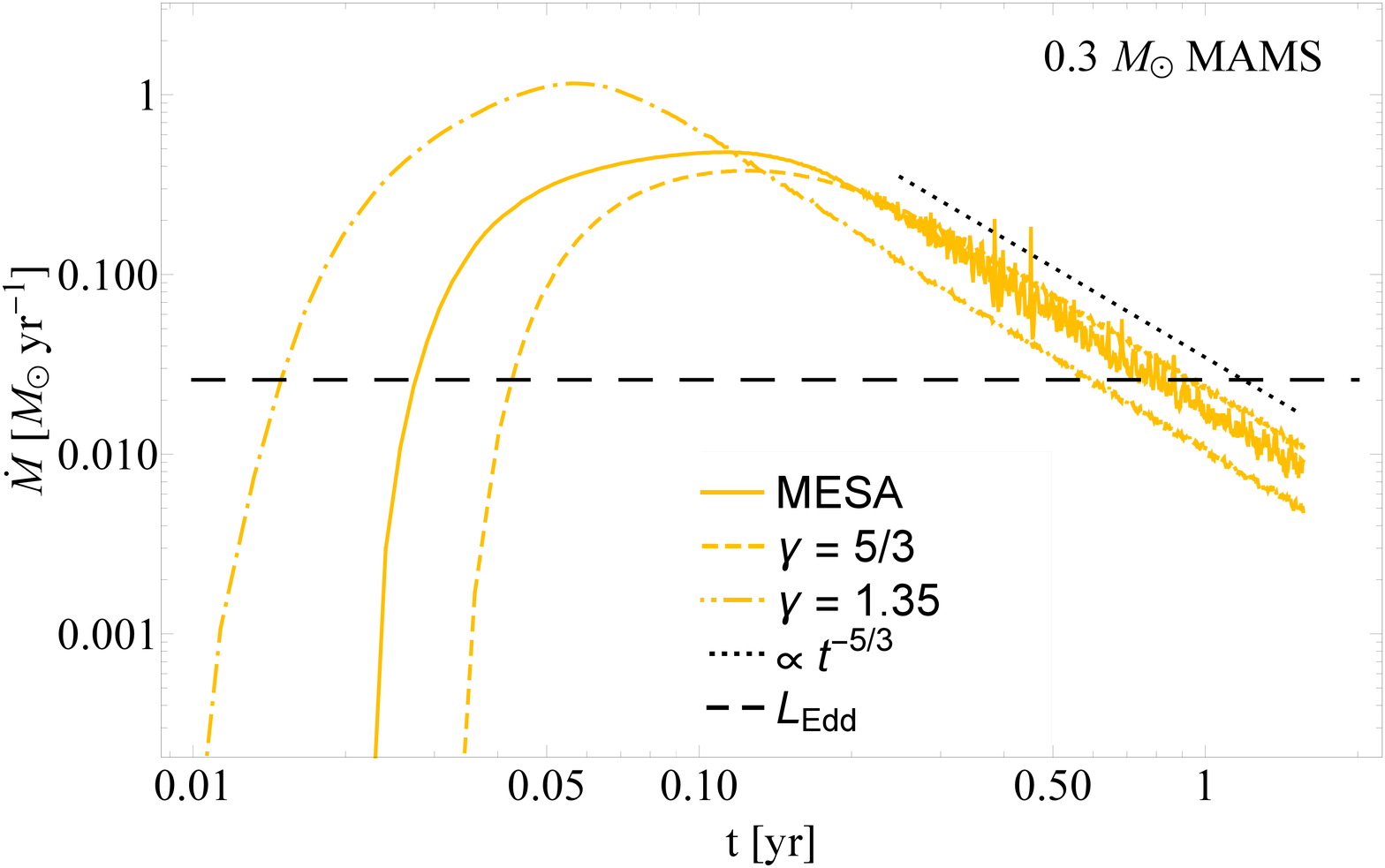}
   \includegraphics[width=0.495\textwidth]{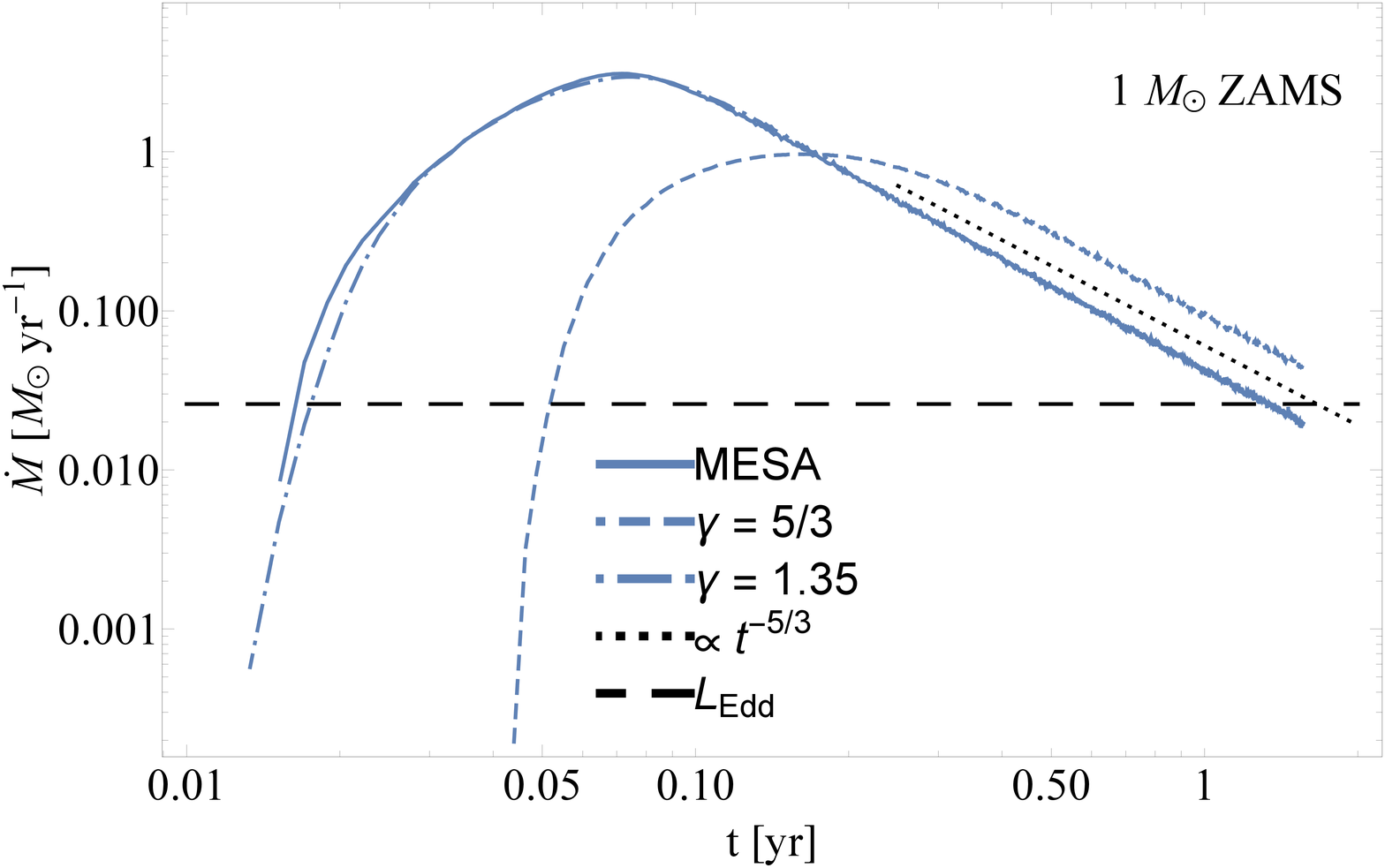}
   \includegraphics[width=0.495\textwidth]{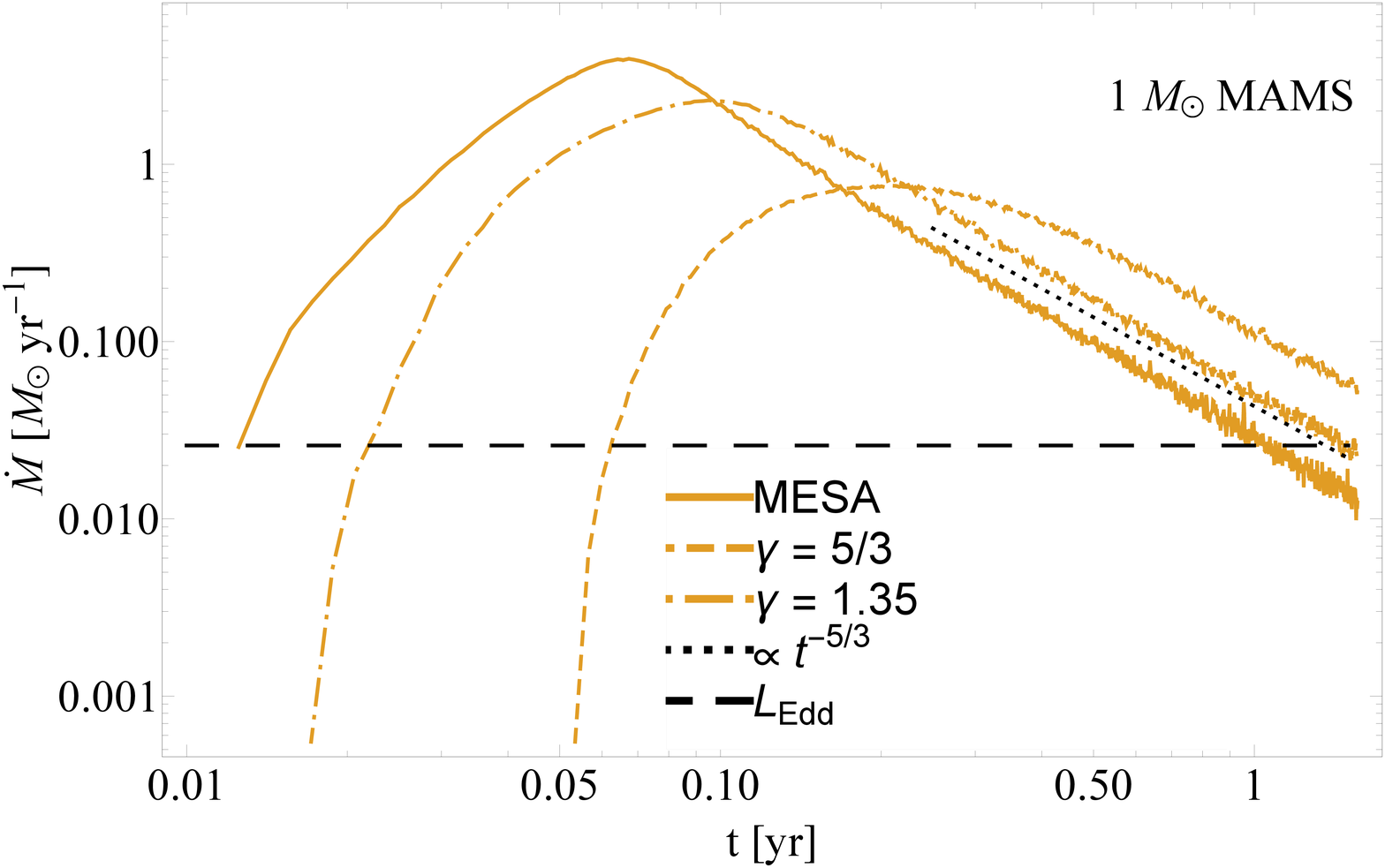}
   \includegraphics[width=0.495\textwidth]{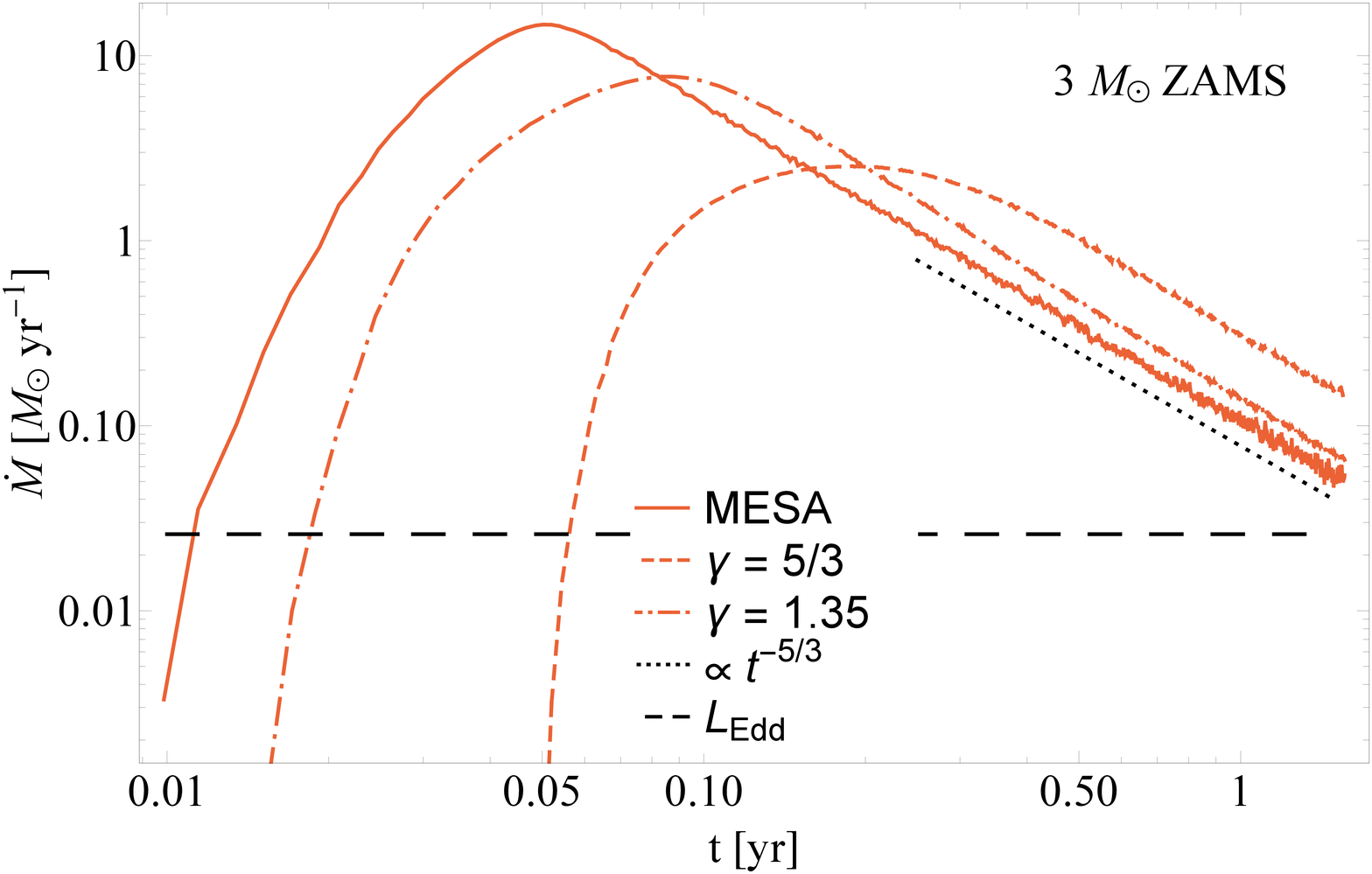}
   \includegraphics[width=0.495\textwidth]{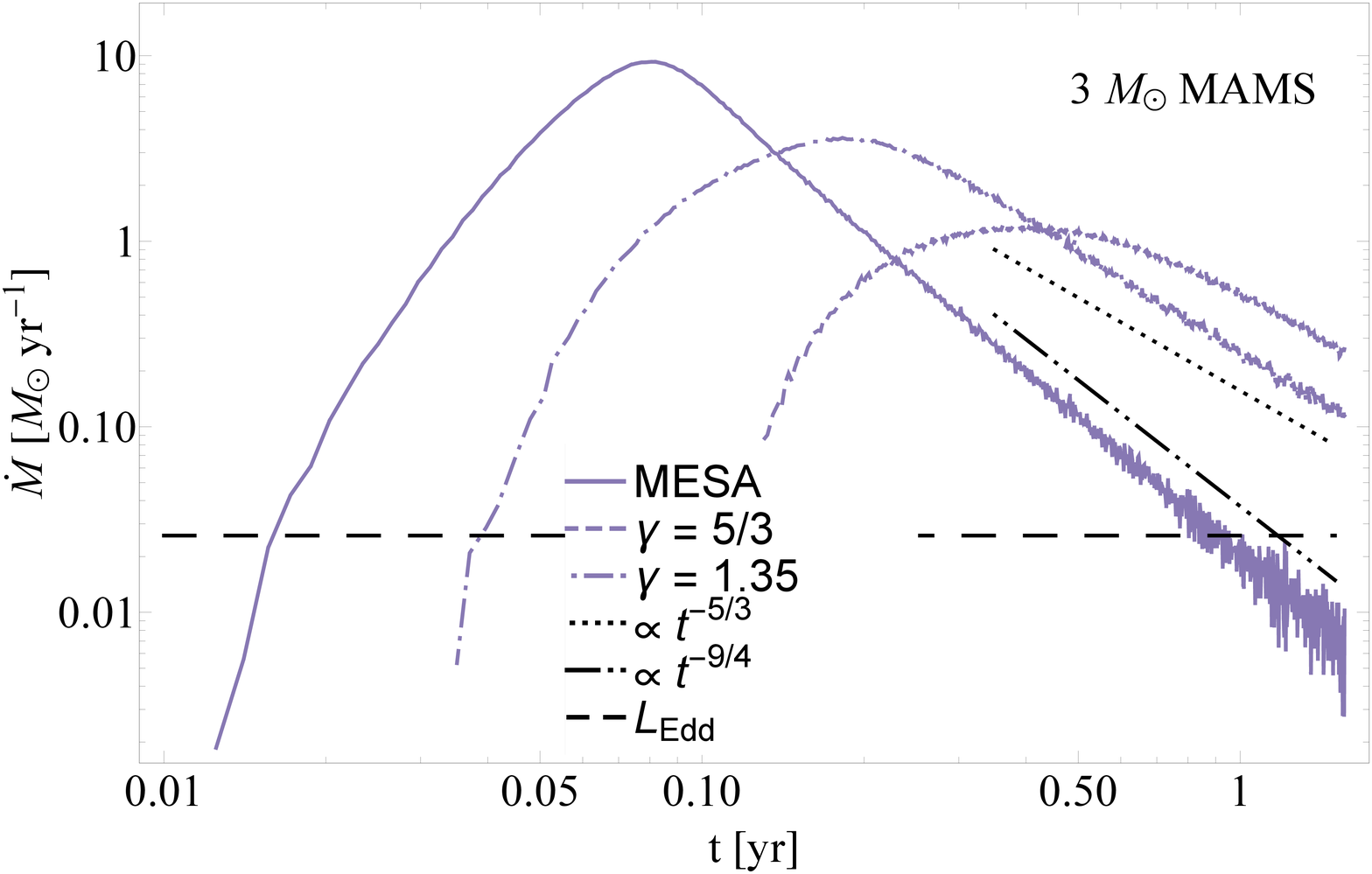}
   \caption{The fallback rate onto the $10^6M_{\odot}$ SMBH in units of Solar masses per year as a function of time in years. Here solid curves correspond to the density profiles generated from {\sc mesa}, dashed curves are $\gamma = 5/3$ polytropes matched to the stellar mass and radius of the {\sc mesa} star, and dot-dashed curves are $\gamma = {{} 1.35}$ polytropes matched to the {\sc mesa} star mass and radius; black, dotted lines show the power-law $\propto t^{-5/3}$, while the dot-dot-dashed line in the bottom-right panel shows the scaling $\propto t^{-9/4}$. The long-dashed, black line gives the Eddington luminosity of the black hole, assuming a radiative efficiency of 10\% and an electron scattering opacity of 0.34 cm$^2$ g$^{-1}$. The specific star is shown by the name in the legend, and panels on the left side show the fallback from stars at the zero-age main sequence, while those on the right are more highly evolved. It is apparent from the top-left and middle-left panels that the fallback curves from the $0.3 M_{\odot}$, ZAMS and the 1$M_{\odot}$, ZAMS progenitors are very well reproduced by $\gamma = 5/3$ and $\gamma = {{} 1.35}$ polytropes, respectively. Every other fallback curve from a {\sc mesa}-generated density profile, however, shows significant deviations from the polytropic approximations. We also see that the $3.0M_{\odot}$, MAMS follows $\propto t^{-9/4}$ at late times, which results from the presence of a bound core that survives the encounter (\citealt{Coughlin:2019aa}; no bound core is left when the star is modeled as a polytrope). The 0.3 $M_{\odot}$, MAMS {\sc mesa} star also shows enhanced variability in the fallback rate, which arises from the fact that the stream -- unlike the polytropic models for the same {\sc mesa} star mass and radius -- has fragmented vigorously into small-scale clumps. }
   \label{fig:fb_zams}
\end{figure*}

To understand the impact of stellar structure, additional simulations were also performed with a $\gamma = 1.35$ polytrope and a $\gamma = 5/3$ polytrope, with the mass and radius of each polytrope matched to those of the {\sc mesa} progenitor and the same orbital properties (i.e., the same $\beta$). For the polytrope disruptions, we set the adiabatic index equal to $\Gamma = 5/3$, such that the microphysics and the stiffness of the equation of state is identical to that employed in the {\sc mesa} calculations. In this way, the only difference between the polytrope disruptions and the {\sc mesa} star disruptions is the density profile -- the stellar mass, radius, and microphysics are identical -- and these simulations therefore isolate the imprint that the density profile has on the fallback rate.

In this paper we are interested in the fallback rate, defined as the rate at which disrupted material returns to pericenter, and we therefore increase the accretion radius to $3\,\,r_{\rm t}$ once the star has passed through pericenter. The fallback rate of material through this radius will, in general, differ from the true accretion rate, which is the rate at which material passes through the horizon of the black hole. In general, the latter requires detailed, high-resolution simulations that accurately model the formation of the accretion flow around the black hole. This is not currently computationally feasible for standard TDE parameters, but has been attempted by various authors in restricted cases (see, e.g., \citealt{hayasaki13, Shiokawa:2015aa, Bonnerot:2016aa, Hayasaki:2016aa, Sadowski:2016aa}). For the relatively high-$\beta$ simulations considered here, the general relativistic advance of periapsis will be large, which should enhance energy dissipation and the formation of an accretion disc. We therefore expect the fallback rates we find to closely track the true accretion rate onto the black hole, but we leave a detailed study of this process to a future investigation (for which our fallback rates could be used as inputs). 

Finally, as we noted above, the more highly evolved {\sc mesa} stars have extremely dense cores, and -- even for a $\beta$ of 3 encounter -- those cores survive the tidal interaction with the black hole. In these instances, the time step is extremely limited because of the high density and sound speed at the center of the stellar remnant, which makes these simulations prohibitively expensive to run for the duration over which the fallback occurs. Therefore, once the surviving core recedes to a {{} significant} distance from the SMBH, we replace all of the particles in the bound core that have a density above the maximum (non-core) stream density by a single sink particle; the position and velocity of the sink is set equal to the center of mass position and velocity of all the particles used to create the sink, and the accretion radius of the sink particle is equal to the maximum distance of these particles from their center of mass (the sink position). {{} For the simulation with the densest core ($0.3M_\odot$, TAMS) we inserted the sink at a time of 3.5 days post pericentre. The other simulations with bound cores were less computationally expensive and ran to later times. As such we inserted the sink at times (post-pericentre) of 15 days ($1.0M_\odot$, TAMS), 11 days ($3.0M_\odot$, TAMS), 173 days ($1.0M_\odot$, MAMS), and 27 days ($3.0M_\odot$, MAMS)}. We tested the robustness of this approach by changing the time at which we replaced the resolved core with the sink, and found negligible changes in the subsequent fallback rate. 

{}{One way in which this approximation could adversely affect the fallback rate is if we erroneously included the marginally-bound (to the core) radius within the particles that constitute the core. If we were to make this error, the fallback rate onto the black hole would abruptly terminate at a late, but ultimately finite time. We have checked that the mass of the sink particle increases very slightly after its formation, indicating that the marginally-bound radius is indeed outside of the radius of the sink particle.}

\subsection{Results}
\begin{figure}
   \centering
   \includegraphics[width=\columnwidth]{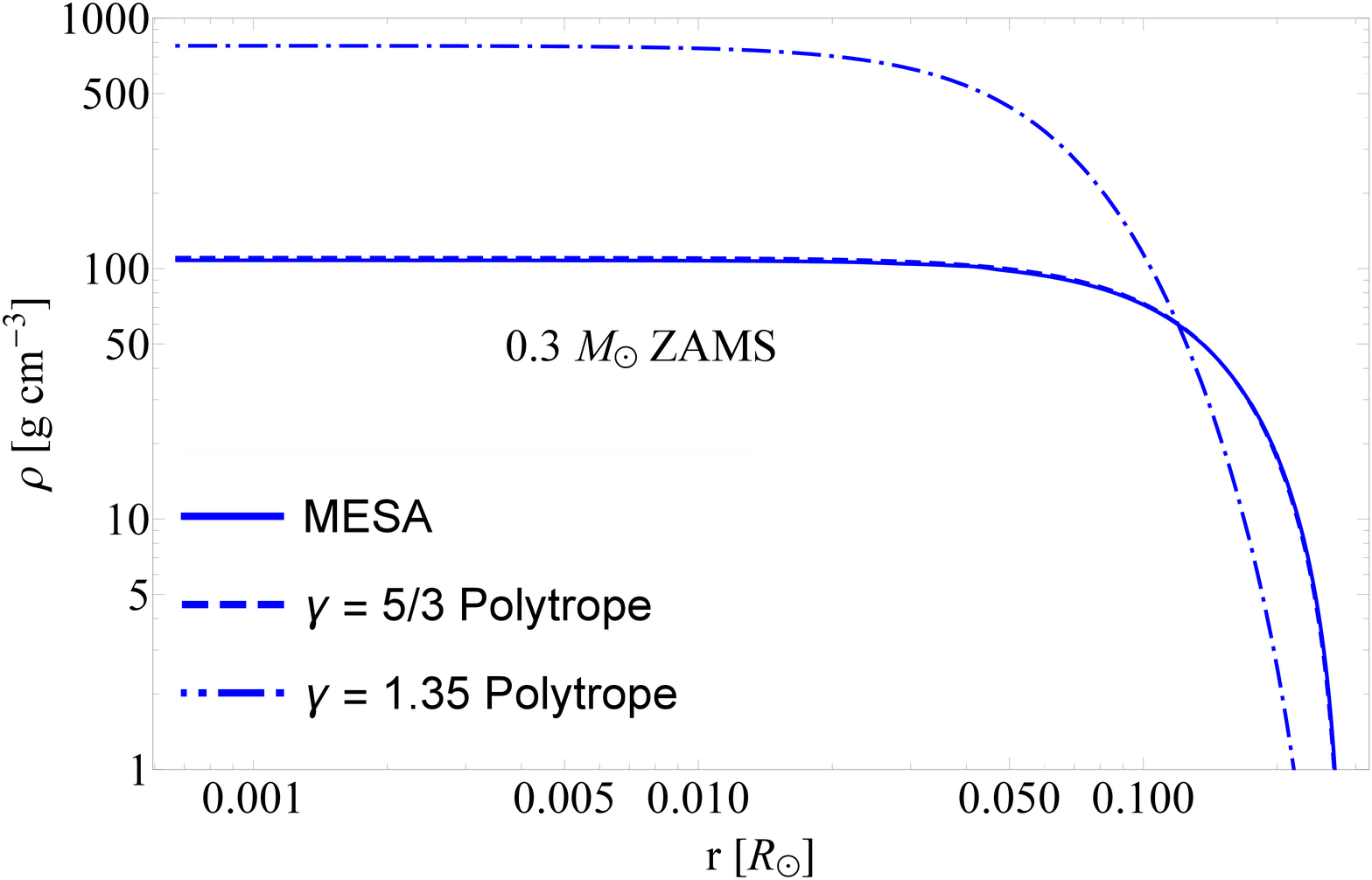} 
   \includegraphics[width=\columnwidth]{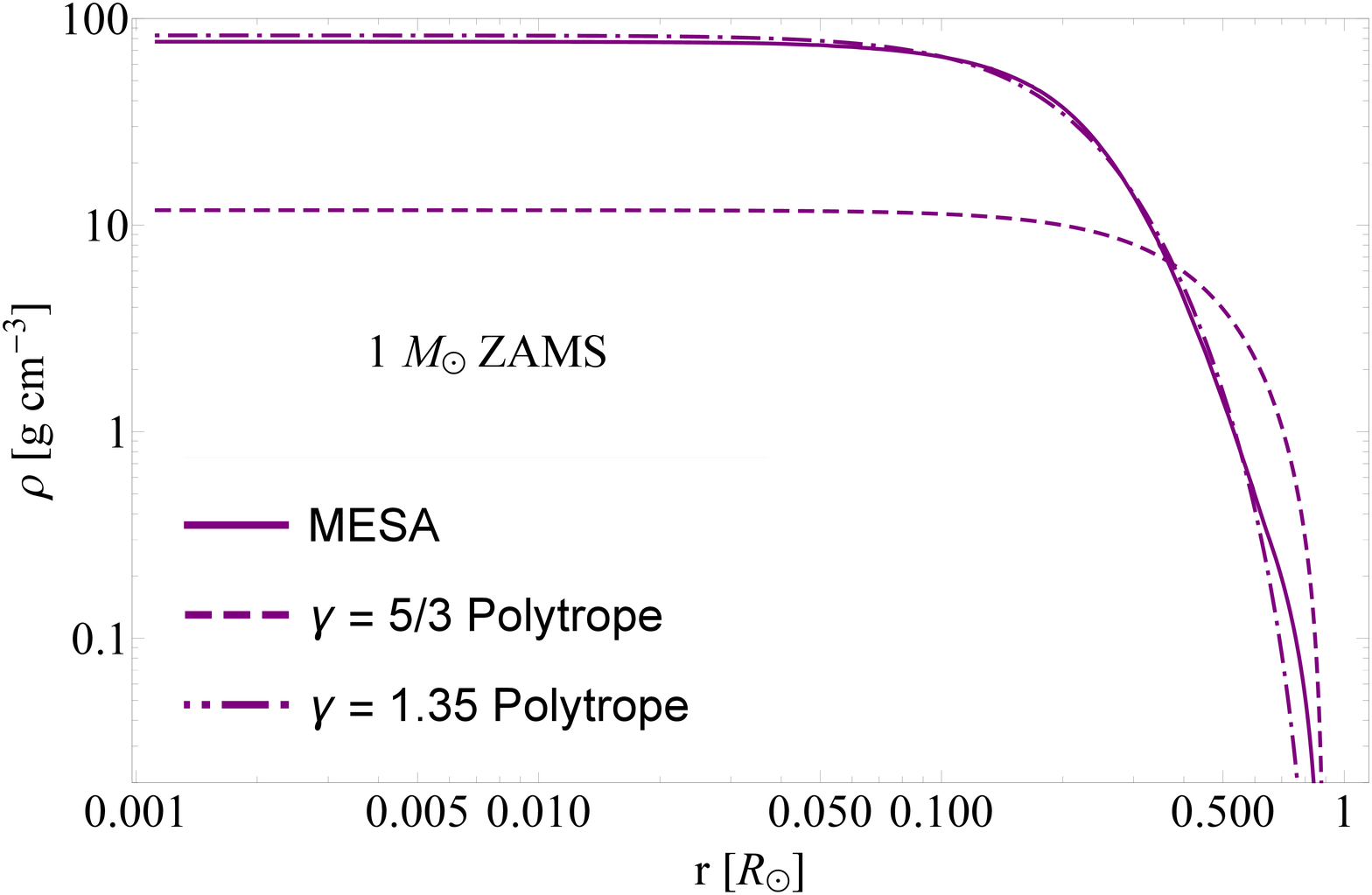} 
   \caption{The density profiles of the $0.3M_\odot$ ZAMS (top) and $1.0M_\odot$ ZAMS (bottom) compared with the density profiles of $\gamma=1.35$ (dot-dashed) and $\gamma=5/3$ (dashed) polytropes with the same mass and radius. For the $0.3M_\odot$ ZAMS star, the $\gamma=5/3$ polytrope provides an excellent fit, while for the $1.0M_\odot$ ZAMS star the $\gamma=1.35$ polytrope provides an excellent fit.}
   \label{fig:lowmass}
\end{figure}

Figure \ref{fig:fb_zams} illustrates the fallback rate onto the black hole in Solar masses per year as a function of time in years from six different stellar models, with the specific stellar model shown in the legend. In each panel the solid curve shows the fallback from the star with the {\sc mesa} density profile, the dashed curve is the $\gamma = 5/3$ model matched to the {\sc mesa} stellar mass and radius, and the dot-dashed curve is the $\gamma = 1.35$ polytropic model (again, with the same mass and radius as the {\sc mesa} star). It is evident from the top-left panel and the middle-left panel that the 0.3 $M_{\odot}$, ZAMS and the 1.0 $M_{\odot}$, ZAMS {\sc mesa} fallback curves are extremely well-reproduced by $\gamma = 5/3$ and $\gamma={{} 1.35}$ polytropes, respectively. This finding indicates, correspondingly, that such stars can be very well-modeled by single polytropes that are gas-pressure and (nearly) radiation-pressure dominated (see Fig.~\ref{fig:lowmass}). This result for the $1M_{\odot}$, ZAMS star was also recovered by \citet{Goicovic:2019aa}, who found that their fallback rates (from a $1M_{\odot}$, ZAMS star generated with {\sc mesa}) were very similar to those of \citet{Guillochon:2013aa}, who used a polytropic approximation.

For every other stellar model, however, there are notable differences between the fallback curves from the {\sc mesa} and the polytropic models. Specifically, we see that employing the {\sc mesa} density profile over either polytropic model systematically shifts the return time of the most-bound debris to earlier times, the time to peak to earlier times, and the magnitude of the peak rate itself is also increased. Furthermore, because the total mass is the same, the larger peak in the accretion rate and the earlier time-to-peak imply that the {\sc mesa} curves must fall below the polytropic ones at some point, and this is indeed recovered in each case. It is also apparent that the {\sc mesa} models conform to a power-law decline at an earlier time than do the polytropic models. 

Every polytropic star is completely disrupted by the SMBH for these $\beta = 3$ encounters, which agrees with the results of \citet{Guillochon:2013aa} and \cite{mainetti17}, who demonstrated that the critical $\beta$ for the full disruption of a $\gamma = 5/3$ polytrope is $\beta \simeq 0.9$, while that for a $\gamma = 4/3$ polytrope is $\beta \simeq 2$. Each ZAMS, {\sc mesa} progenitor is also fully disrupted. However, more highly evolved stars start to yield partial TDEs, and the {\sc mesa} $1M_{\odot}$ MAMS and $3.0M_{\odot}$ MAMS leave stellar cores at the location of the marginally-bound orbit. We see that, for the case of the $3.0M_{\odot}$, MAMS progenitor, the presence of the core has the affect of modifying the late-time fallback rate, which declines approximately as $\propto t^{-9/4}$ \emph{instead of $t^{-5/3}$}. This result is in agreement with the analytic predictions of \citet{Coughlin:2019aa}.

\begin{figure*}[t!]
   \centering
   \includegraphics[width=0.495\textwidth]{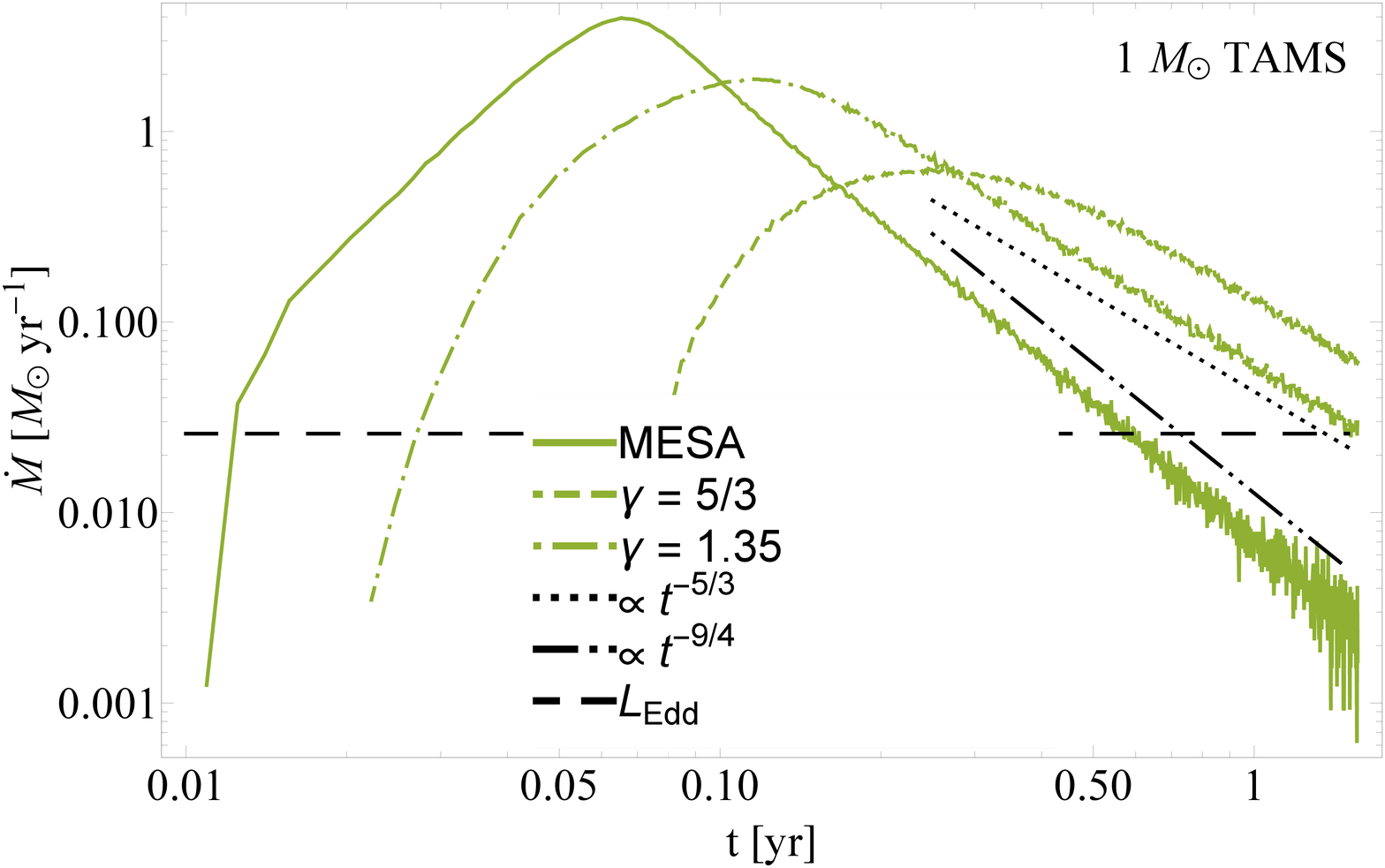} 
   \includegraphics[width=0.495\textwidth]{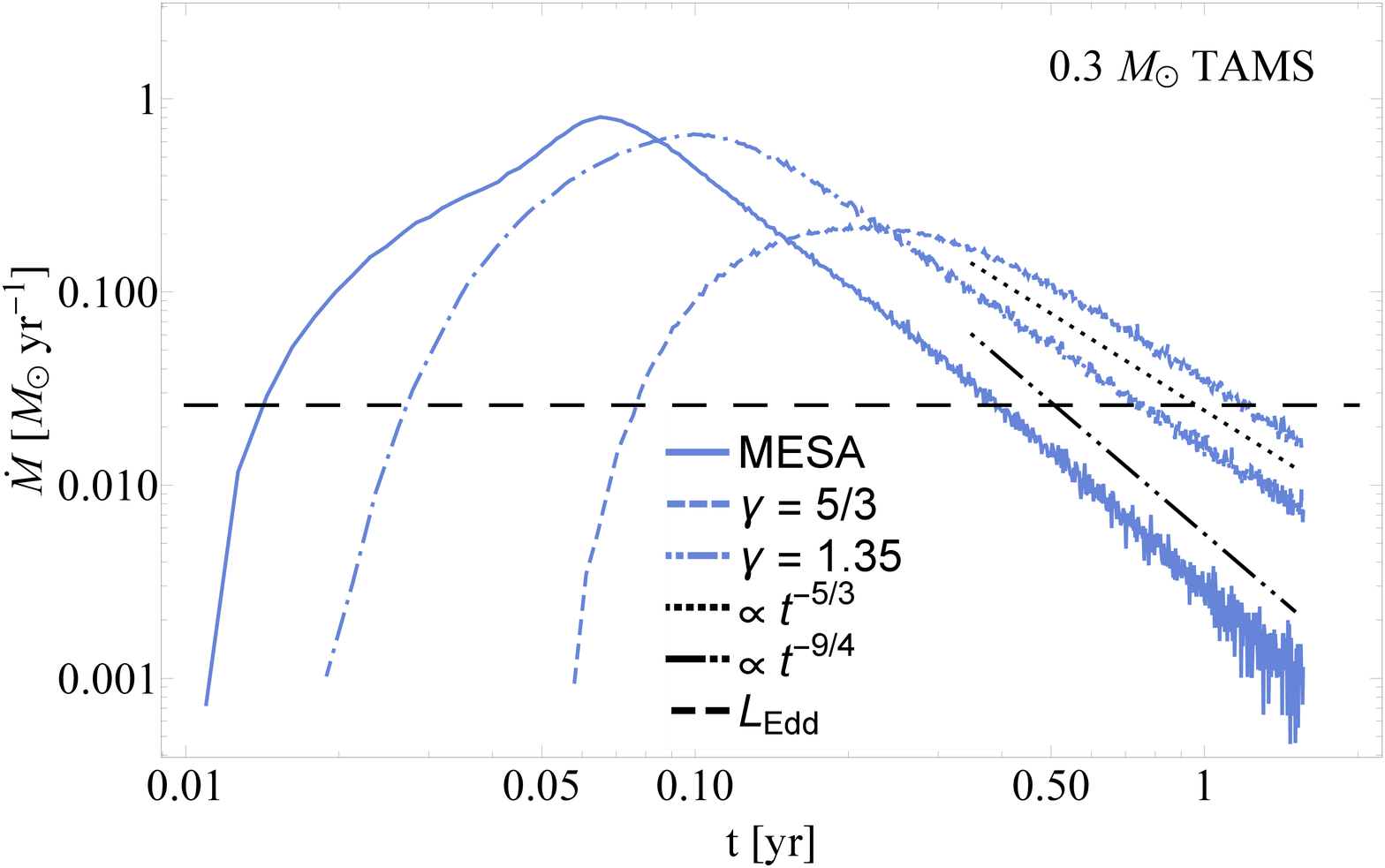} 
   \caption{The fallback rate from the $1M_{\odot}$, TAMS progenitor (left) and the $0.3 M_{\odot}$, TAMS progenitor (right), where the solid curves are for the {\sc mesa}-generated density profile and dashed (dot-dashed) curves are from $\gamma = 5/3$ ($\gamma = {{} 1.35}$) polytropes matched to the {\sc mesa} star mass and radius. In each case the {\sc mesa} density profile yields a bound core that survives the encounter, while the polytropes do not, which results in a late-time power-law falloff that declines approximately as $t^{-9/4}$ (dot-dot-dashed line) and is significantly steeper than $t^{-5/3}$ (dotted line). The Eddington luminosity of the black hole -- assuming a radiative efficiency of 10\% and an electron-scattering-dominated opacity of 0.34 cm$^2$ g$^{-1}$ -- is shown by the long-dashed, black line. }
   \label{fig:fb_tams}
\end{figure*}

The TAMS, {\sc mesa} progenitors all possess extremely dense cores that survive the tidal encounter, each of which modifies the late-time fallback rate onto the black hole, as shown in Figure \ref{fig:fb_tams} (the time taken for the $3.0M_{\odot}$ progenitor to go from MAMS to TAMS is very short, and hence the TAMS fallback curve appears nearly identical to the bottom-right panel of Figure \ref{fig:fb_zams}; we therefore opted not to show this fallback rate). We also emphasize that the lifetime of the $0.3 M_{\odot}$ star is well in excess of the age of the Universe, and hence this star cannot be disrupted by a SMBH (at least not any time soon). However, the \emph{density profile} of this star could conceivably be achieved by a more massive progenitor -- which would evolve to the TAMS within the age of the Universe -- with different initial properties (e.g., metallicity, rotation).

\begin{figure}[htbp]
   \centering
   \includegraphics[width=\columnwidth]{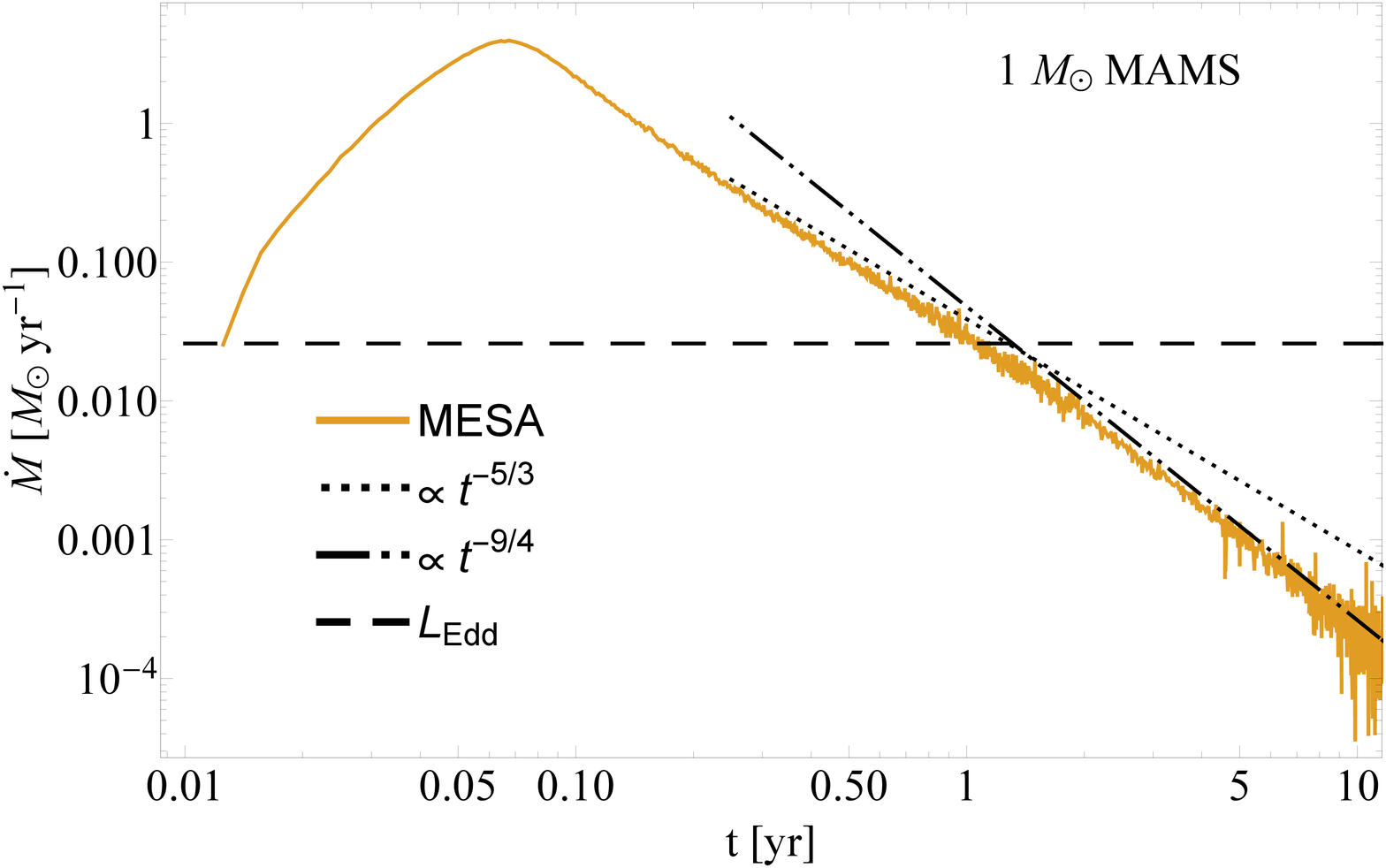} 
   \caption{The fallback rate from the $1M_{\odot}$, MAMS {\sc mesa} progenitor, in units of Solar masses per year as a function of time in years, run out to 10 years post-disruption. The dotted line shows the scaling $\propto t^{-5/3}$ (predicted to be the asymptotic power-law followed by the fallback if there is no surviving core), the dot-dot-dashed line gives the scaling $\propto t^{-9/4}$ (predicted by \citealt{Coughlin:2019aa} to be the asymptotic power-law decline of the fallback if there is a surviving core), and the long-dashed line gives the Eddington luminosity of the hole if the radiative efficiency is $10\%$ and the opacity is set to the electron scattering opacity of $0.34$ cm$^2$ g$^{-1}$. Here the stream possesses a gravitationally-bound core (with a mass of $\sim 15\%$ of the progenitor star) at the location of the marginally-bound radius that reforms out of the stream after the star is initially completely disrupted. We see that for roughly the first year after the return of the most-bound debris there is little evidence of the existence of the core on the fallback, and the fallback curve appears to asymptote to a $t^{-5/3}$ decline. However, around roughly 1 year, there is a noticeable break in the falloff, and the curve steepens to a decline that is well-matched by the power-law $\propto t^{-9/4}$.}
   \label{fig:mams_fblong}
\end{figure}

As we noted above, the $1M_{\odot}$, MAMS, {\sc mesa} progenitor also possesses a bound stellar core at the center of mass of the tidally-disrupted debris stream. In this case, however, the star is \emph{initially completely disrupted}, and the core recollapses out of the stream at a time significantly after the stellar center of mass passes through pericenter. For this reason, the mass of the surviving core is only a small fraction of the initial progenitor ($\simeq 15\%$), and consequently the fallback curve shows little evidence of the gravitational influence of the core over $\sim 1$ yr and appears to approach a decline $\propto t^{-5/3}$.

Figure \ref{fig:mams_fblong} shows the fallback from the $1M_{\odot}$, MAMS, {\sc mesa} progenitor out to 10 years post-disruption and demonstrates, however, that the presence of the core does start to affect the fallback at later times. In particular, we see that while the first year shows little evidence of the existence of a bound core, there is a clear break in the fallback curve at a time of 1-2 years where the power-law of the fallback rate transitions from $\propto t^{-5/3}$ to one that is better matched by $\propto t^{-9/4}$. Interestingly, this time at which a break in the power-law is exhibited is very close to the time predicted by the analytical model in \citet{Coughlin:2019aa} (see their Figure 2), and {coincidentally} also occurs around the same time at which the fallback rate drops below the Eddington limit of the SMBH (black, dashed line, assuming a radiative efficiency of 10\% an{}d an electron-scattering opacity of 0.34 cm$^{2}$ g$^{-1}$).

\section{Implications for Black Hole Mass Estimates}
\label{sec:massestimate}
It is apparent from Figures \ref{fig:fb_zams} and \ref{fig:fb_tams} that, for the majority of progenitors, non-polytropic stellar structure generates substantial differences in the fallback rate onto the black hole. Notably, the {\sc mesa} profiles yield earlier times to peak and larger peak fallback rates, and they more rapidly approach a power-law falloff as compared to the polytropes. To use these differences to estimate the corresponding differences in the inferred black hole mass that would arise by assuming a given density profile, we must have a mapping between a characteristic timescale (e.g., the time to peak), the black hole mass, and the properties of the star. As shown in Section \ref{sec:impulse}, when the fallback rate is computed with the impulse approximation, this mapping arises through the timescale

\begin{equation}
T_{\rm mb} \simeq \frac{2\pi R_{\star}^{3/2}M}{M_{\star}\sqrt{GM}} \propto M^{1/2}, \label{Tmb}
\end{equation}
which is the return time of the most-bound debris. Additional dependence on the stellar structure modifies the fallback rate through a dimensionless function of time normalized by $T_{\rm mb}$, and that dimensionless function can be calculated from the (assumed-unaltered) density profile when the star is at the tidal radius. Thus, the dependence on the black hole mass arises only as $\propto M^{1/2}$, and \emph{in this approximation}, fallback curves are simply scaled in time and magnitude by $\sqrt{M}$. 

By comparing Figure \ref{fig:mdots1} to Figures \ref{fig:fb_zams} and \ref{fig:fb_tams}, wee see that the frozen-in approximation does not accurately reproduce many of the features of the numerically-obtained fallback rates. In addition to the fact that the time to peak is shorter and the peak itself is higher in the numerical simulations (by a factor $\gtrsim 10$), the ordering of the curves is actually \emph{inverted} between the two approaches: while Figure \ref{fig:mdots1} shows that the $\gamma = 5/3$ polytrope peaks earlier than $\gamma = 1.35$ polytrope, which itself peaks earlier than the {\sc mesa} model, Figures \ref{fig:fb_zams} and \ref{fig:fb_tams} demonstrate that the $\gamma = 5/3$ polytrope always reaches a peak \emph{after} the $\gamma = 1.35$ polytrope\footnote{This effect can also be seen in Figures 2 and 10 of \citet{Lodato:2009aa}, though this inversion was not noted by those authors.}. Moreover, for every case except the $0.3 M_{\odot}$ ZAMS and $0.3M_{\odot}$ MAMS progenitors, the {\sc mesa} model peaks earlier than the $\gamma = 1.35$ polytrope. The numerically-obtained return time of the most bound debris also differs for each density profile, whereas, under the frozen-in approximation, this timescale -- for the same black hole mass -- is only affected by the stellar mass and radius (which, for a given star, are identical by construction).

These discrepancies indicate that the impulse approximation does not include enough physics to accurately capture the bulk features of the fallback rate. As discussed at length in \citet{Coughlin:2016aa}, it is likely that the most crucial physical ingredient lacking from the impulse approximation is the self-gravity of the debris stream, as the stellar center of mass rapidly recedes outside of the tidal sphere of the black hole. At this point, the stellar density is comparable to the ``black hole density'', being $\rho_{\bullet} \sim M/r_{\rm t}^3$, and the self-gravity of the stream is capable of competing against the shear of the black hole. The self-gravity of the stream induces density waves that traverse the stream radially, and these waves serve to generate more pronounced ``shoulders'' near the extremities of the stream and flatten the $dm/dr \propto \rho$ curve from the polytropic one that follows from the frozen-in approximation. It is, in fact, because of this nearly-flat $dm/dr$ generated by self-gravity that the fallback curves more rapidly approach the $t^{-5/3}$ decline (or the $t^{-9/4}$ decline for the partial TDEs). The higher central density of the $\gamma = 1.35$ polytrope also generates more vigorous density waves, which correspondingly produce a flatter density distribution and give rise to an earlier time-to-peak as compared to a $\gamma = 5/3$ polytrope. 

Nonetheless, it is likely that after some amount of time following the disruption, the mass distribution is \emph{approximately} frozen-in, meaning that self-gravity has smoothed out any density perturbations and the stream is long enough that the time-dependent potential due to self-gravity is small\footnote{However, the arguments of \citet{Coughlin:2016aa} and the simulations of \citet{Coughlin:2015aa} suggest that the stream is weakly gravitationally unstable, and hence the freezing of the energy distribution is only valid over an integrated region of the stream that contains many clumps that form out of the instability.}. In this case, the energies of gas parcels comprising the stream are still Keplerian in the potential of the black hole, and the energies themselves are simply established at some later time; indeed, these arguments were used -- and verified by a direct evaluation of the energy distribution at different times -- by \citet{Lodato:2009aa} and \citet{Guillochon:2013aa} to calculate fallback rates to the black hole after only a small fraction of the return time of the most bound debris had been directly simulated\footnote{{}{When the tidal encounter leaves a surviving core behind, a Keplerian energy distribution is no longer upheld; however, as shown by \citet{Coughlin:2019aa}, when the self-gravity of the stream itself no longer significantly modifies the density distribution along the stream, one can make a change of variables when calculating the fallback rate that shows that Equation \eqref{tc}, and hence Equation \eqref{M2M1_2}, still holds.}}. Moreover, if for a given $\beta$ and stellar progenitor the density profile at the time the energy is frozen-in is \emph{independent of the black hole mass}, which the simulations of \citet{Wu:2018aa} verify\footnote{This is also a reasonable expectation, as $\beta$ measures the tidal strength of the black hole; thus, for encounters with the same $\beta$, it follows that the energy distribution should be roughly fixed at the same time after self-gravity (which depends only on the stellar properties) has modified the density distribution, and the absolute value of the black hole mass should not matter. This assumption breaks down, however, once the orbital timescale becomes shorter than the time over which self-gravity acts to modify the density distribution, which occurs for very small black hole masses (where even the tidal approximation itself starts to break down).} (see their Figure 1), then it follows that any physical timescale in a TDE can be written

\begin{equation}
t_{\rm c} = \sqrt{M}f_{\star,\beta}, \label{tc}
\end{equation}
where $f_{\star,\beta}$ is a function that depends only on the stellar properties and $\beta$. We therefore see that, for the same star and the same orbital parameters, we recover the same result as we did in Section \ref{sec:impulse}: for two TDEs with identical orbital and stellar properties and physical timescales $t_1$ and $t_2$, we can satisfy $t_1 = t_2$ by changing the black hole mass $M_1$ to $M_2$, with $M_2$ given by

\begin{equation}
M_2 = M_1\left(\frac{t_2}{t_1}\right)^2. \label{M2M1_2}
\end{equation}

As an example, the $3M_{\odot}$, MAMS progenitor (bottom-right panel of Figure \ref{fig:fb_zams}) modeled as a $\gamma = 1.35$ polytrope has a time to go from the first half max, $t_{\rm half, 1}$, to max, $t_{\rm max}$, of $t_1 = t_{\rm max}-t_{\rm half,1} \simeq 0.085$ yr. On the other hand, the {\sc mesa} model of the same star has $t_2 = t_{\rm max} - t_{\rm half,1} = 0.0275$ yr. Thus, \emph{if we modeled the disruption of the {\sc mesa} star as a polytrope}, then we would require a black hole mass of $M_2 = (t_2/t_1)^2M_1 \simeq 0.097M_1 \simeq 10^5M_{\odot}$ to reproduce the observed timescale. 

Table \ref{tab:2} gives the ratio $M_2/M_1$ required to shift the timescale of the polytropic star to the timescale reproduced by the {\sc mesa}-star disruption. The timescale itself is shown in the top row of the table, where $t_{\rm max} - t_{\rm half,1}$ is the time to go from first-half-max to the peak fallback rate, $t_{\rm half,2} - t_{\rm max}$ is the time taken to fall by a factor of two below the peak fallback rate, and $t_{\rm half,2} - t_{\rm half,1}$ is the full width at half maximum of the fallback curve. The stellar progenitor is given in the left-most column of the table, and the value in the left (right) of each cell is the ratio $M_2/M_1$ required to yield the {\sc mesa}-generated timescale by modeling the star as a $\gamma = 1.35$ ($\gamma = 5/3$) polytrope. For example, if we were to model the $0.3M_{\odot}$ star as a $\gamma = 1.35$ polytrope, then we would require a black hole mass of $M_2 = 5\times M_1$ to reproduce the timescale $t_{\rm max} -t_{\rm half,1}$ found from the disruption of the {\sc mesa} model, and the number $M_2/M_1 = 5$ is shown in the top-left cell of the table.

\begin{figure}[htbp] 
   \centering
   \includegraphics[width=0.495\textwidth]{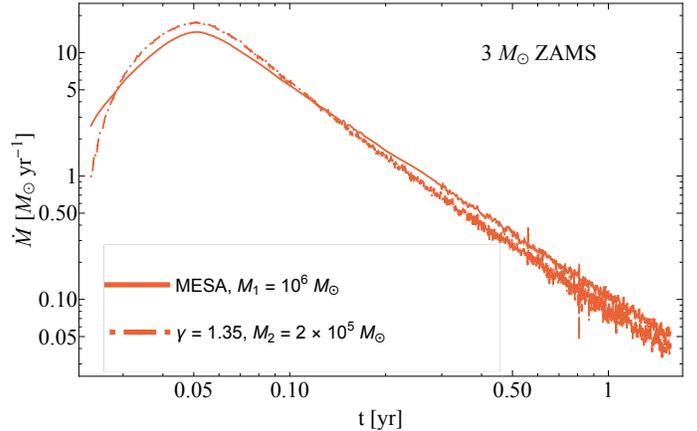} 
   \caption{{{} The fallback curve from the $3M_{\odot}$, ZAMS {\sc mesa} progenitor, with a black hole mass of $M_{1} = 10^6M_{\odot}$ (red, solid; this curve is identical to the solid, red curve shown in the bottom-left panel of Figure \ref{fig:fb_zams}). The dot-dashed curve shows the fallback rate from the tidal disruption of a $\gamma = 1.35$ polytrope, with its mass and radius matched to those of the {\sc mesa} star. Here, however, we reduced the mass of the SMBH by the factor shown in Table \ref{tab:2}, corresponding to a black hole mass of $M_{2} = 2\times 10^{5}M_{\odot}$, and we aligned the first time to half max to that of the {\sc mesa} fallback curve (i.e., we first ``see'' the TDE at the first-half-max). This figure demonstrates that we can reproduce the ``data'' obtained from the disruption of the {\sc mesa} star extremely well with a polytrope, but with a black hole mass that differs by nearly an order of magnitude from the true value.}}
   \label{fig:3MSun_BHChanged}
\end{figure}

We see from this table that, for stellar density profiles that are well-reproduced by polytropes (the $0.3M_{\odot}$ ZAMS and the $1M_{\odot}$ ZAMS stars, as shown in Fig.~\ref{fig:lowmass}), the mass ratio required to reproduce the {\sc mesa}-generated timescale with a polytropic one is very close to unity. In these instances, the inferred black hole mass estimate is not far off the true, underlying value. However, for more massive progenitors and more highly evolved stars, the extremely dense core of the {\sc mesa} progenitor shifts each timescale earlier, which consequently requires a significantly smaller black hole mass to yield the same characteristic timescale with a polytropic model.

{}{As a direct demonstration of this effect, Figure \ref{fig:3MSun_BHChanged} illustrates the fallback rate from the disruption of the $3M_{\odot}$, ZAMS {\sc mesa} model by a SMBH of mass $M_1 = 10^6M_{\odot}$, which is the same curve shown in the bottom-left panel of Figure \ref{fig:fb_zams}. The dot-dashed curve is the fallback curve from the disruption of a $\gamma = 1.35$ polytrope, the stellar mass and radius identical to those of the {\sc mesa} star; in this case, however, we set the mass of the disrupting SMBH to $M_{2} = 2\times 10^5$, which is -- from Table \ref{tab:2} -- the value predicted to equate the time to go from the first half max to the peak between the two models, and we aligned the first time to half max of the polytrope fallback curve to that of the {\sc mesa} star (i.e., we initially ``see'' both disruptions at the same time, that time being the first time to half max). We see that this polytropic model provides an extremely good fit to the ``data'' obtained from the fallback curve of the {\sc mesa} model, but at the expense of incorrectly inferring the SMBH mass \emph{by nearly an order of magnitude}.}

Of course, our approach here to ``modeling'' the lightcurve of a TDE is overly simplistic, as one does not necessarily have any prior information about the nature of the progenitor or the black hole, and one must use a combination of timescales to recover the best-fit model parameters (as is done in, for example, \citealt{guillochon18}). However, these results do demonstrate that care needs to be taken to ensure that the template fallback curves used to interpret observed data sets contain a sufficiently broad range of stellar density profiles (e.g., accounting for stellar age, and non-polytropic density profiles) to ensure that the inferred parameters are accurate and that the error bars that result from the data fitting are appropriate.

\begin{table*}
\scriptsize
\begin{center}
\begin{tabular}{|l|c|c|c|} \hline
\diagbox[dir =SE, outerleftsep=-.18cm, innerleftsep=2.5cm, innerwidth=-.4cm, innerrightsep=.01cm, outerrightsep=-.28cm]{\hspace{-3.2cm} Star}{Timescale}&\hspace{-0.6cm} \begin{tabular}{c}$t_{\rm max}-t_{\rm half,1}$ \\ \hspace{-1.1cm} \begin{tabular}{ccc}$\gamma = 1.35$ & \, & $\gamma = 5/3$\end{tabular}\end{tabular} & \hspace{-0.6cm} \begin{tabular}{c}$t_{\rm half,2}-t_{\rm max}$ \\ \hspace{-1.1cm} \begin{tabular}{cc}$\gamma = 1.35$ & $\gamma = 5/3$\end{tabular}\end{tabular} & \hspace{-0.6cm} \begin{tabular}{c}$t_{\rm half,2}-t_{\rm half,1}$ \\ \hspace{-1.1cm} \begin{tabular}{cc}$\gamma = 1.35$ & $\gamma = 5/3$\end{tabular}\end{tabular} \\
\hline
$0.3M_{\odot}$ ZAMS &\hspace{-.8cm} \begin{tabular}{cc} {\scriptsize $M_2/M_1= 5.0$} &  {\scriptsize $M_2/M_1= 0.81$ } \end{tabular} & \hspace{-.8cm} \begin{tabular}{cc} {\scriptsize $M_2/M_1= 12$} &  {\scriptsize $M_2/M_1= 1.19$ } \end{tabular} &  \hspace{-.8cm} \begin{tabular}{cc} {\scriptsize $M_2/M_1= 9.1$} &  {\scriptsize $M_2/M_1= 1.07$ } \end{tabular} \\
\hline
$0.3M_{\odot}$ MAMS &\hspace{-.8cm} \begin{tabular}{cc} {\scriptsize $M_2/M_1= 7.0$} &  {\scriptsize $M_2/M_1= 1.2$ } \end{tabular} & \hspace{-.8cm} \begin{tabular}{cc} {\scriptsize $M_2/M_1= 5.3$} &  {\scriptsize $M_2/M_1= 0.50$ } \end{tabular} &  \hspace{-.8cm} \begin{tabular}{cc} {\scriptsize $M_2/M_1= 5.9$} &  {\scriptsize $M_2/M_1= 0.67$ } \end{tabular} \\
\hline
$0.3M_{\odot}$ TAMS &\hspace{-.8cm} \begin{tabular}{cc} {\scriptsize $M_2/M_1= 0.24$} &  {\scriptsize $M_2/M_1= 0.061$ } \end{tabular} & \hspace{-.8cm} \begin{tabular}{cc} {\scriptsize $M_2/M_1= 0.22$} &  {\scriptsize $M_2/M_1= 0.022$ } \end{tabular} &  \hspace{-.8cm} \begin{tabular}{cc} {\scriptsize $M_2/M_1= 0.23$} &  {\scriptsize $M_2/M_1= 0.030$ } \end{tabular} \\
\hline
$1.0M_{\odot}$ ZAMS &\hspace{-.8cm} \begin{tabular}{cc} {\scriptsize $M_2/M_1= 0.83$} &  {\scriptsize $M_2/M_1= 0.15$ } \end{tabular} & \hspace{-.8cm} \begin{tabular}{cc} {\scriptsize $M_2/M_1= 0.92$} &  {\scriptsize $M_2/M_1= 0.074$ } \end{tabular} &  \hspace{-.8cm} \begin{tabular}{cc} {\scriptsize $M_2/M_1= 0.89$} &  {\scriptsize $M_2/M_1= 0.093$ } \end{tabular} \\
\hline
$1.0M_{\odot}$ MAMS &\hspace{-.8cm} \begin{tabular}{cc} {\scriptsize $M_2/M_1= 0.33$} &  {\scriptsize $M_2/M_1= 0.042$ } \end{tabular} & \hspace{-.8cm} \begin{tabular}{cc} {\scriptsize $M_2/M_1= 0.25$} &  {\scriptsize $M_2/M_1= 0.025$ } \end{tabular} &  \hspace{-.8cm} \begin{tabular}{cc} {\scriptsize $M_2/M_1= 0.27$} &  {\scriptsize $M_2/M_1= 0.031$ } \end{tabular} \\
\hline
$1.0M_{\odot}$ TAMS &\hspace{-.8cm} \begin{tabular}{cc} {\scriptsize $M_2/M_1= 0.14$} &  {\scriptsize $M_2/M_1= 0.032$ } \end{tabular} & \hspace{-.8cm} \begin{tabular}{cc} {\scriptsize $M_2/M_1= 0.13$} &  {\scriptsize $M_2/M_1= 0.010$ } \end{tabular} &  \hspace{-.8cm} \begin{tabular}{cc} {\scriptsize $M_2/M_1= 0.14$} &  {\scriptsize $M_2/M_1= 0.015$ } \end{tabular} \\
\hline
$3.0M_{\odot}$ ZAMS &\hspace{-.8cm} \begin{tabular}{cc} {\scriptsize $M_2/M_1= 0.19$} &  {\scriptsize $M_2/M_1= 0.033$ } \end{tabular} & \hspace{-.8cm} \begin{tabular}{cc} {\scriptsize $M_2/M_1= 0.22$} &  {\scriptsize $M_2/M_1= 0.019$ } \end{tabular} &  \hspace{-.8cm} \begin{tabular}{cc} {\scriptsize $M_2/M_1= 0.21$} &  {\scriptsize $M_2/M_1= 0.023$ } \end{tabular} \\
\hline
$3.0M_{\odot}$ MAMS &\hspace{-.8cm} \begin{tabular}{cc} {\scriptsize $M_2/M_1= 0.097$} &  {\scriptsize $M_2/M_1= 0.018$ } \end{tabular} & \hspace{-.8cm} \begin{tabular}{cc} {\scriptsize $M_2/M_1= 0.059$} &  {\scriptsize $M_2/M_1= 0.0057$ } \end{tabular} &  \hspace{-.8cm} \begin{tabular}{cc} {\scriptsize $M_2/M_1= 0.072$} &  {\scriptsize $M_2/M_1= 0.0086$ } \end{tabular} \\
\hline
$3.0M_{\odot}$ TAMS &\hspace{-.8cm} \begin{tabular}{cc} {\scriptsize $M_2/M_1= 0.067$} &  {\scriptsize $M_2/M_1= 0.0076$ } \end{tabular} & \hspace{-.8cm} \begin{tabular}{cc} {\scriptsize $M_2/M_1= 0.045$} &  {\scriptsize $M_2/M_1= 0.0049$ } \end{tabular} &  \hspace{-.8cm} \begin{tabular}{cc} {\scriptsize $M_2/M_1= 0.053$} &  {\scriptsize $M_2/M_1= 0.0058$ } \end{tabular} \\
\hline
\end{tabular}
\caption{The ratio $M_2 / M_1$ that is required to produce the same physical timescale if the {\sc mesa} fallback curve is modeled by a polytropic one. Here the physical timescale {}{used to infer the required mass ratio} is given in the top row, with $t_{\rm max} - t_{\rm half, 1}$ the time from the first-half-max to the maximum fallback rate, $t_{\rm half,2} - t_{\rm max}$ the time taken to go from the peak fallback rate to half that value, and $t_{\rm half,2} - t_{\rm half,1}$ the full-width at half-maximum{}{; each one of these timescales differs for a given star, but by scaling the black hole mass by the value shown in each cell, they can be brought into agreement with one another}. The stellar model is given in the left-most column, and the ratio $M_2/M_1$ obtained by using a $\gamma = 1.35$ ($\gamma = 5/3$) is given in the left (right) of each cell.}
\label{tab:2}
\end{center}
\end{table*}

\section{Discussions and conclusions}
\label{sec:conclusions}
In this paper we presented simulations of the tidal disruption of stars encountering an SMBH. We modeled the tidally-disrupted stars with the same bulk properties (mass and radius) using three different prescriptions: (1) a $\gamma = 5/3$, polytropic density profile, (2) a $\gamma = 1.35$ $(\approx 4/3)$, polytropic density profile, and (3) a density profile calculated from the {\sc mesa} stellar evolution code. Our simulated disruptions included stars with masses of $0.3M_\odot$, $1.0M_\odot$ and $3.0M_\odot$, each of which was evolved to the zero-age main sequence, the terminal-age main sequence (where the hydrogen mass fraction in the core fell below 0.1\%), and the ``middle age main sequence'', which we defined to be the time at which the hydrogen mass fraction in the core fell below 20\%. We therefore simulated a total of 27 disruptions (9 different stars, each star modeled with a {\sc mesa} density profile and two different polytropic profiles).

In each simulation we maintained the same physics: we employed a polytropic equation of state where the Lagrangian entropy was fixed and set to ensure the isolated star was in hydrostatic balance (cf.~Figure \ref{fig:density}); we fixed the adiabatic index in each simulation to $5/3$, which ensures that the dynamics of the stream's self-gravity \citep[cf.][]{Coughlin:2015aa} differs only by the mass-entropy distribution along the stream; and we fixed the tidal effects from the black hole on each star by employing a $\beta \equiv r_{\rm t}/r_{\rm p} = 3$ for each simulation, where $r_{\rm p}$ ($r_{\rm t}$) is the pericenter (tidal) radius of the star (see also Table \ref{tab:mesa_properties}). Our simulations therefore isolate the impact of the stellar density profile calculated from {\sc mesa} when compared to those calculated by $\gamma = 5/3$ and $\gamma = 1.35$ polytropes.

In general we find that there are significant differences in the simulated fallback rates for stars with different masses and different ages, and further that in most cases these fallback rates deviate significantly from predictions made using polytropes. The exceptions, which come as no surprise as their structures are accurately modelled by polytropes (see Figure \ref{fig:lowmass}), are the $0.3M_\odot$ ZAMS star -- which is well-modelled by a $\gamma=5/3$ polytrope -- and the $1.0M_\odot$ ZAMS star -- which is well-modelled by a $\gamma={{} 1.35}$ polytrope. At both MAMS and TAMS we find that neither polytrope provides an acceptable description for the fallback curve. Similarly the fallback rates from the $3.0M_\odot$ stars are all significantly different to the fallback rates from either polytrope.

There are also differences found in the overall dynamics of the disruption event. In several cases, most notably for the $0.3M_\odot$ MAMS star, the debris stream is significantly more self-gravitating for the {\sc mesa} star than for the polytropes. This results in the fallback curve exhibiting more variability on the power-law decay (see the top-right panel of Fig.~\ref{fig:fb_zams}). We also find that several of the {\sc mesa} stars are not fully disrupted, even with an impact parameter of $\beta=3$, and this arises from the more centrally-concentrated nature of the {\sc mesa} stars\footnote{This finding implies that the classical tidal radius, which depends on the average stellar density, is only an indicator of the distance at which tides from the black hole become important for the majority of the star {by volume}, and not necessarily the core.}. The difficulty of fully disrupting real stars, and particularly more highly-evolved stars, implies that a greater fraction of the events we observe will be partial, rather than full, disruptions (though we note that stars spend more of their lives near the zero-age main sequence, where the {\sc mesa} profiles still yield full disruptions for $\beta=3$). It has recently been shown \citep{Coughlin:2019aa} that TDEs that leave a bound core have a fallback rate whose power-law index asymptotes to $\approx -9/4$ rather than the usual $-5/3$. In each case that leaves a bound core, our simulations recover this result. In future work we will explore whether simulations are also capable of recovering, e.g., the time at which the power-law slope changes to this value as a function of the mass of the core that survives the encounter. We find \citep[see also][]{Guillochon:2013aa} that in some cases, here for the $1.0M_\odot$ MAMS {\sc mesa} star, that the core can be initially fully disrupted but reform after leaving the tidal radius. We attribute this to the velocity field imparted in the stellar debris by the black hole tides, which can at later times cause the stream to converge along its width and augment the density within the stream \citep{Coughlin:2016aa, steinberg19}.

In Section~\ref{sec:massestimate} we described the impact of using realistic stellar models in simulations of TDEs on the inference of the black hole mass from observed TDE lightcurves. We showed that for many types of stars, and particularly those that are more massive at the zero-age main sequence or more highly evolved, the shape of the fallback curves from simulations that employ polytropic stellar models can lead to large errors in the estimated black hole mass. Therefore, employing accurate models for the imprint of stellar structure on the fallback rate to produce templates for TDE lightcurves appears essential for accurately inferring the black hole mass. It could also be that degeneracies between the parameters in TDEs (e.g., stellar mass, stellar age, stellar spin, stellar metallicity, black hole mass, black hole binarity, black hole spin, impact parameter, inclination and phase angles of the stellar orbit, and the accretion dynamics on small scales) mean that an unambiguous estimation of system parameters (or at least an understanding of the true level of error within those estimates) requires accurate and detailed modeling of the stars that are being disrupted.

\acknowledgments
{}{We thank the anonymous referee for a useful report.} CJN is supported by the Science and Technology Facilities Council (grant number ST/M005917/1). ERC acknowledges support from the Lyman Spitzer Jr.~Postdoctoral Fellowship, and from NASA through the Einstein Fellowship Program, grant PF6-170170, and the Hubble Fellowship, grant \#HST-HF2-51433.001-A awarded by the Space Telescope Science Institute, which is operated by the Association of Universities for Research in Astronomy, Incorporated, under NASA contract NAS5-26555. 

\software{{\sc splash}\citep{Price:2007aa}, {\sc phantom} \citep{Price:2018aa}, \sc{mesa} \citep{Paxton:2011aa,Paxton:2013aa,Paxton:2015aa,Paxton:2018aa}}

\appendix
\section{Stellar structure with {\sc phantom}}
\begin{figure*}
   \centering
   \includegraphics[width=0.3\textwidth]{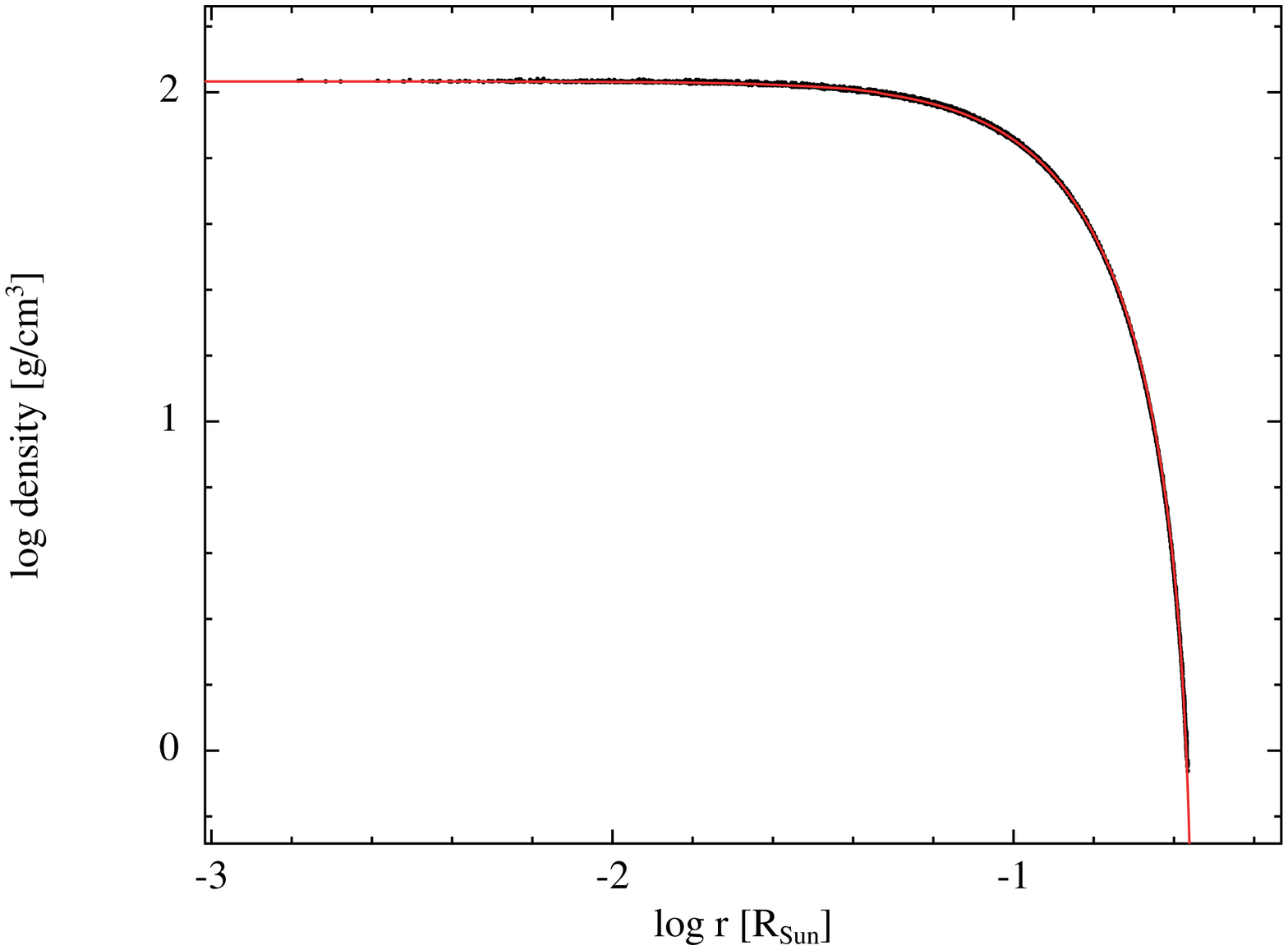}\hfill 
   \includegraphics[width=0.3\textwidth]{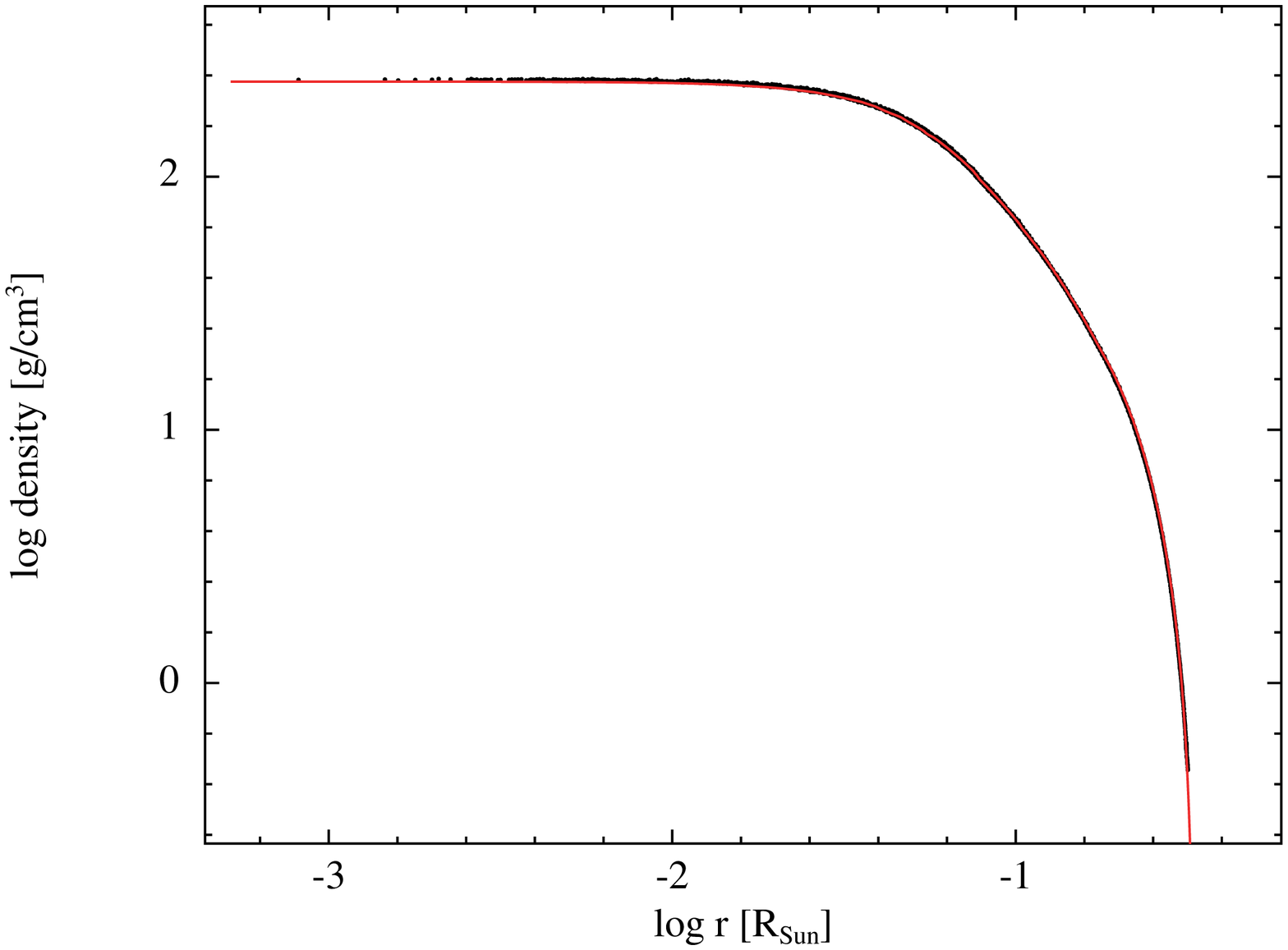}\hfill
   \includegraphics[width=0.3\textwidth]{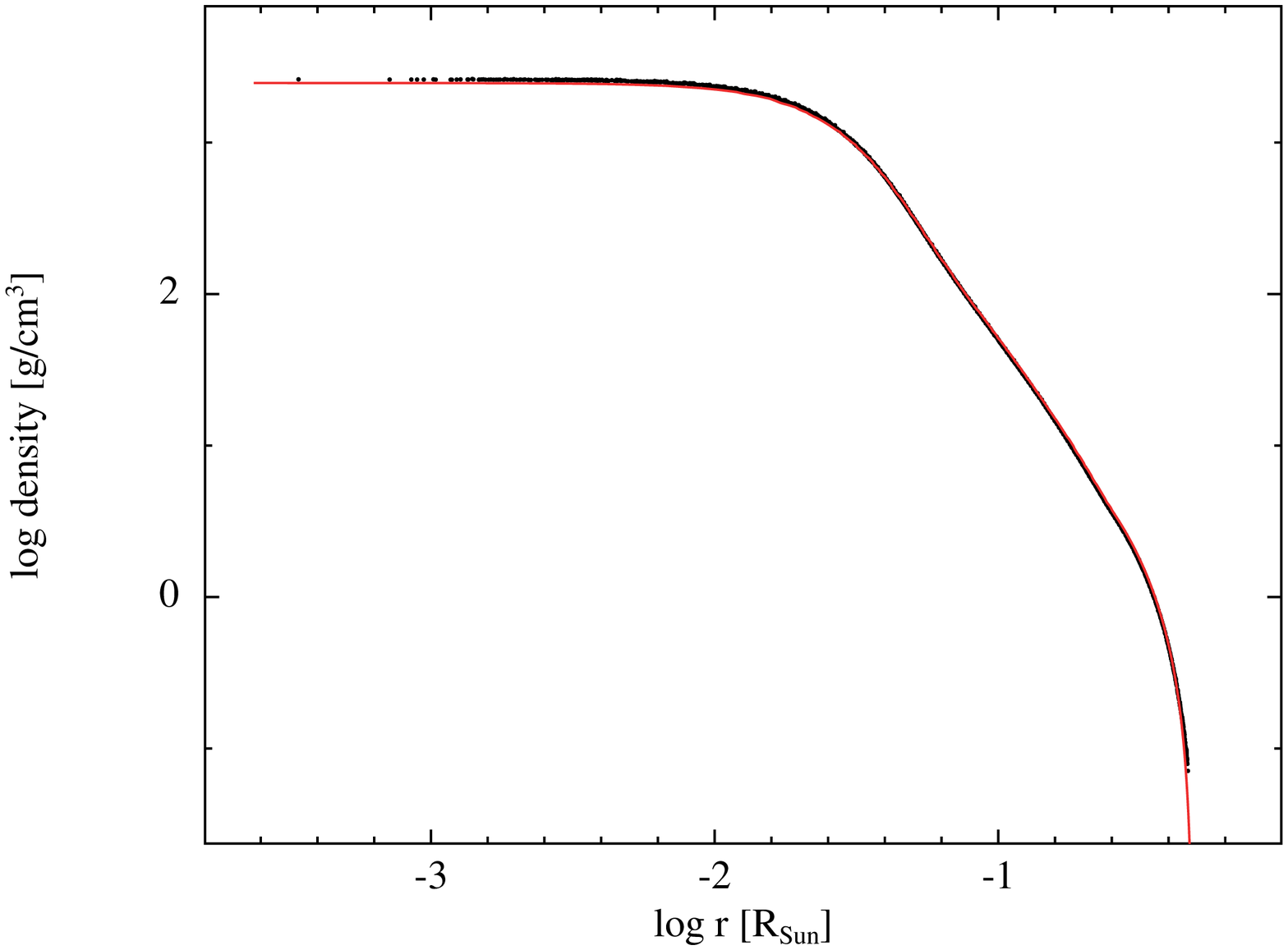}
   \includegraphics[width=0.3\textwidth]{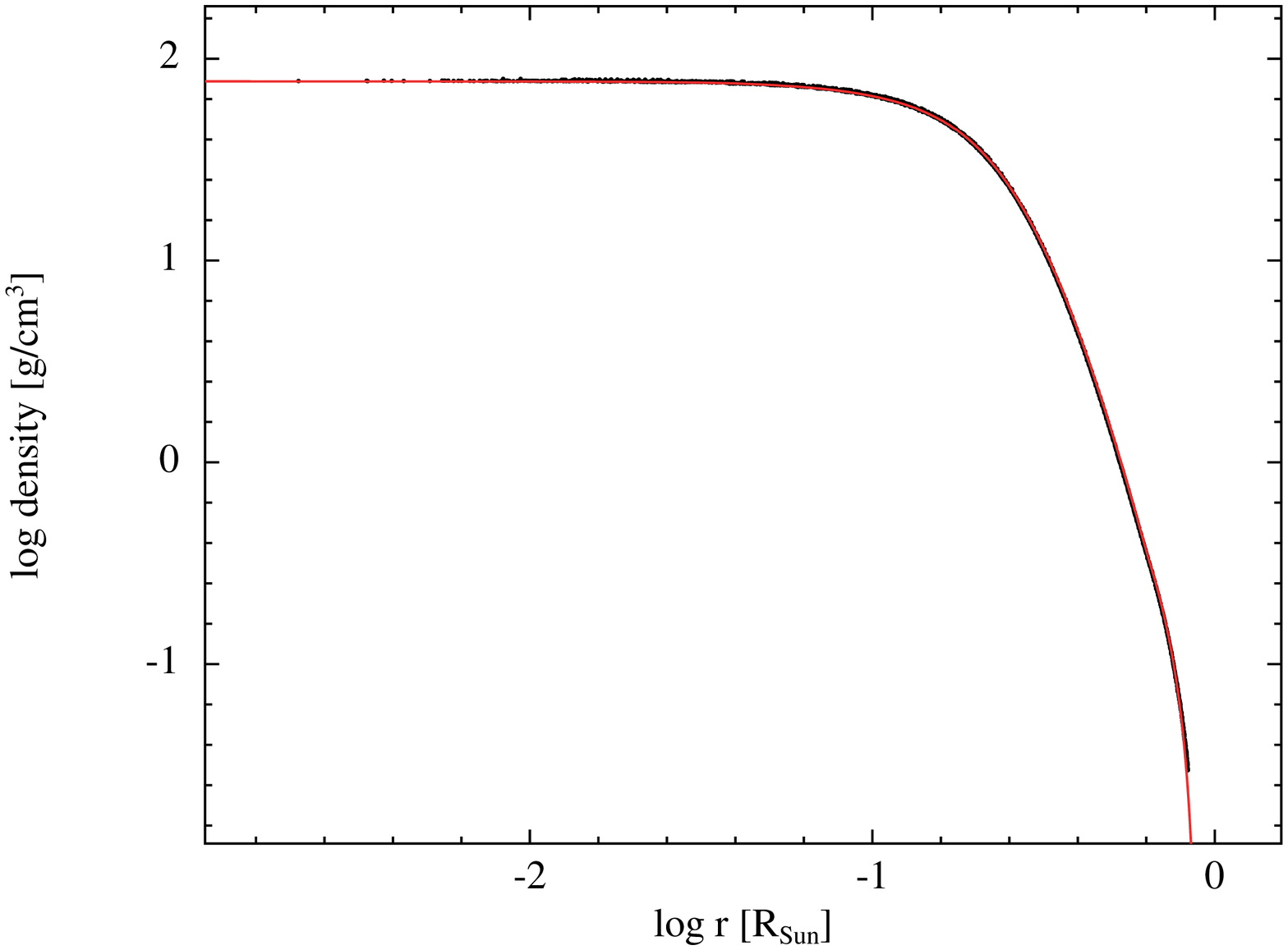}\hfill
   \includegraphics[width=0.3\textwidth]{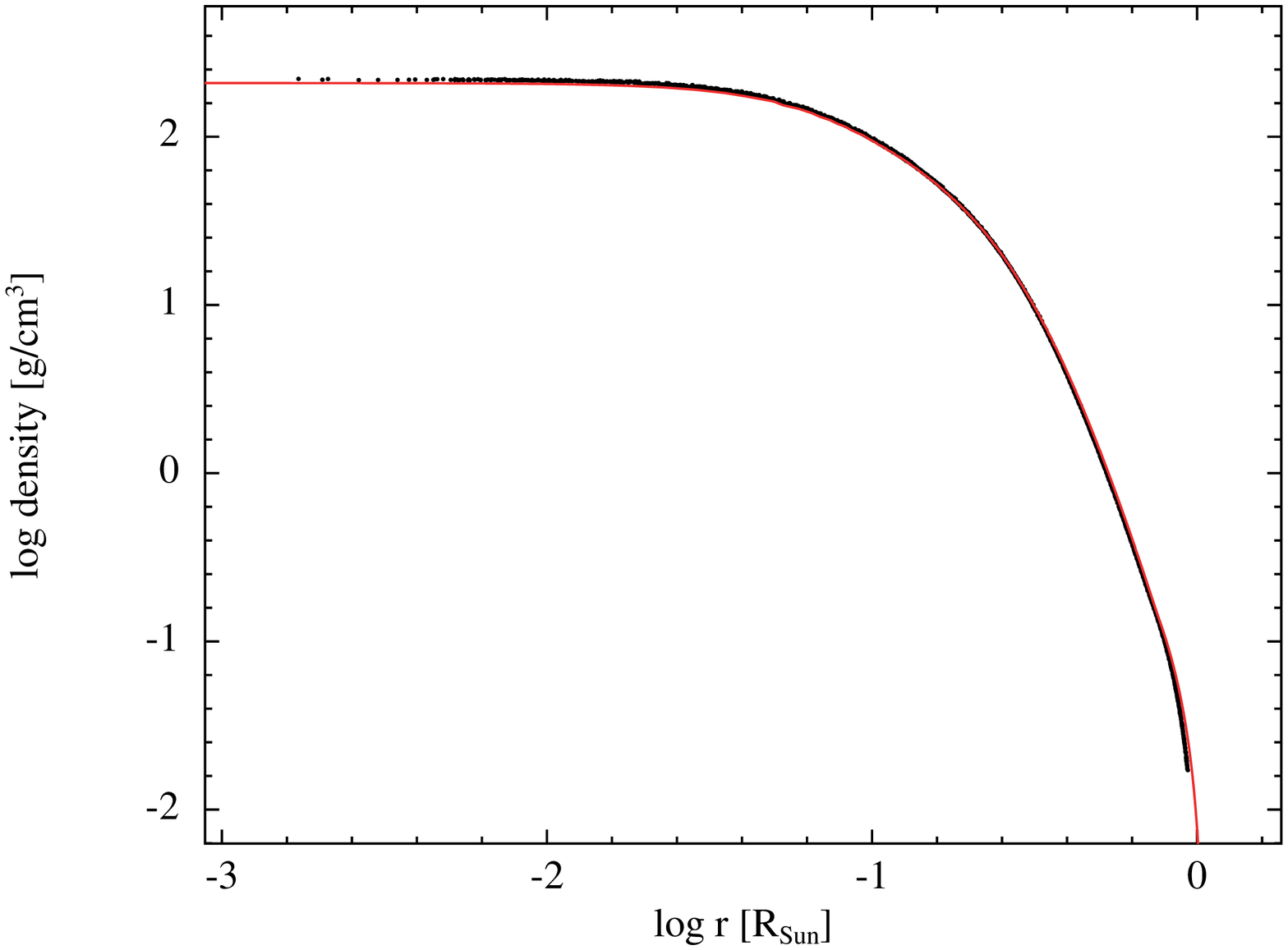}\hfill
   \includegraphics[width=0.3\textwidth]{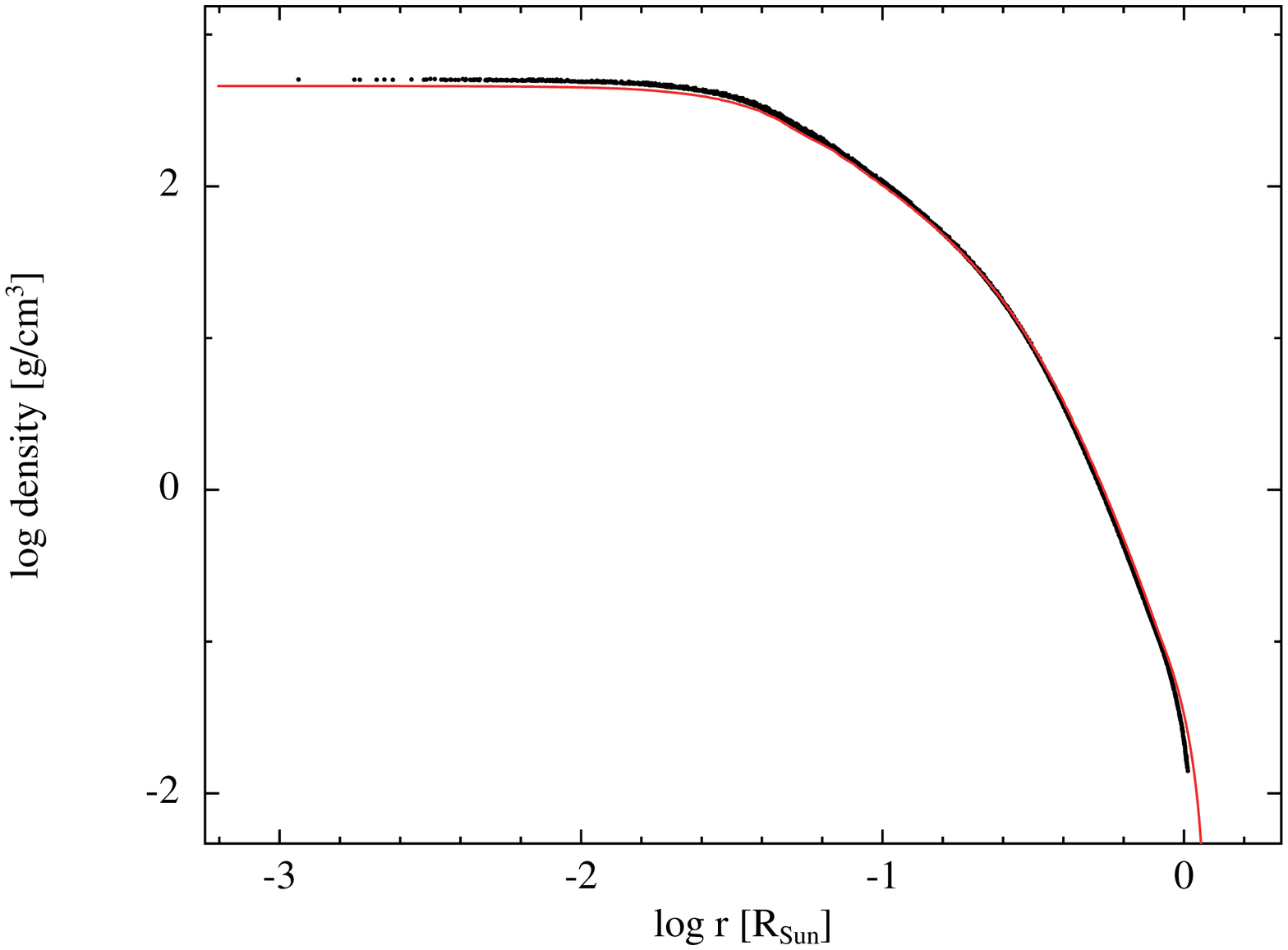} 
   \includegraphics[width=0.3\textwidth]{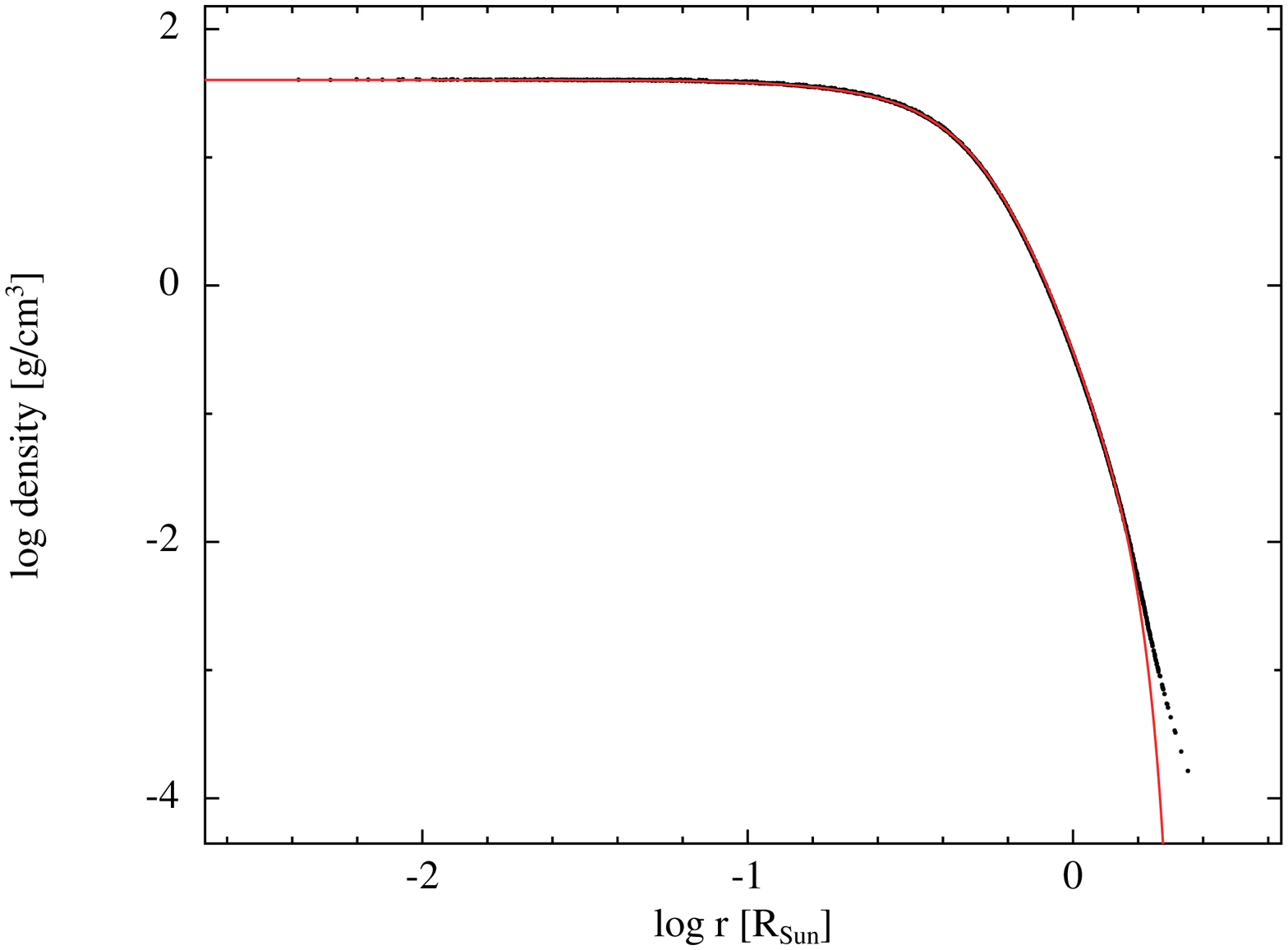}\hfill
   \includegraphics[width=0.3\textwidth]{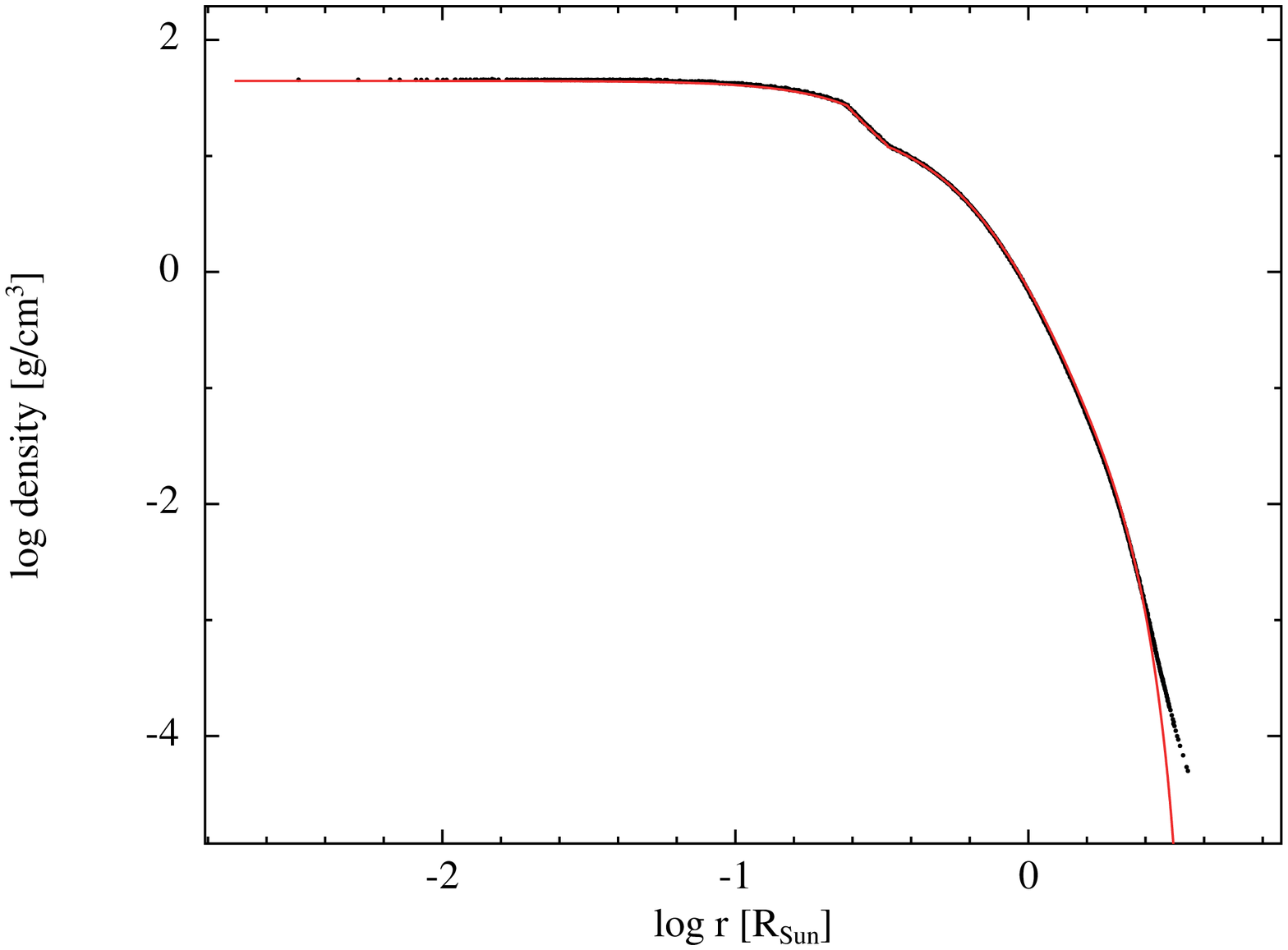}\hfill
   \includegraphics[width=0.3\textwidth]{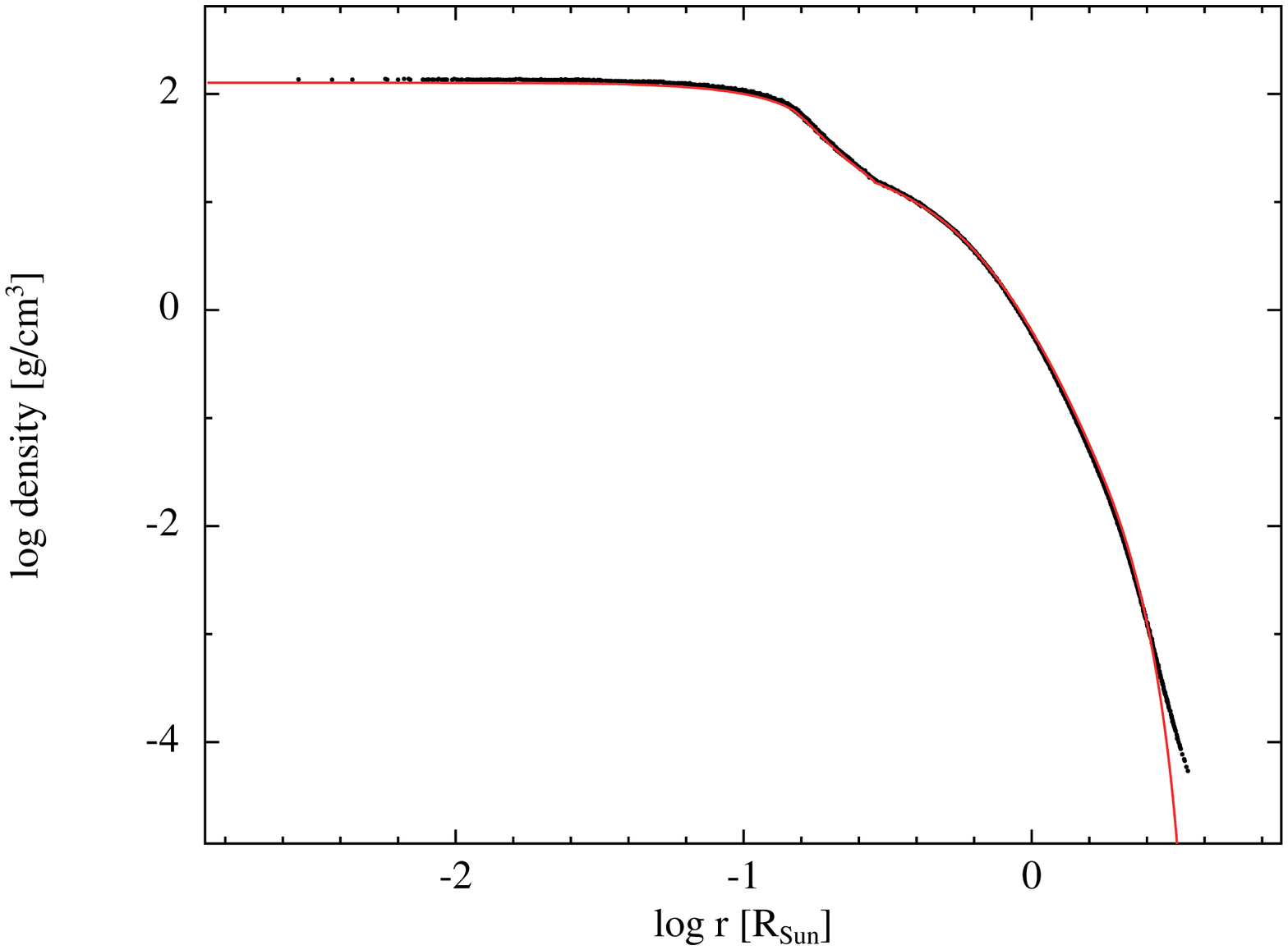} 
   \caption{Initial relaxed density profiles of stars with {\sc mesa} structures. Top row is for $M_\star = 0.3M_\odot$, middle row is $M_\star = M_\odot$ and bottom row is $M = 3.0M_\odot$. Left hand column is at ZAMS, middle column is MAMS and right hand column is TAMS.}
   \label{fig:density}
\end{figure*}

Figure \ref{fig:density} illustrates a comparison between the density profiles obtained with {\sc mesa} (red curves) and those obtained with {\sc phantom} after the initial particle distribution has been relaxed (black curves). For the $1M_{\odot}$, TAMS star, the {\sc phantom} profile overshoots the {\sc mesa} one by about 10\%, and the density of the very outermost radii of the 3 $M_{\odot}$ star (at all ages) is slightly larger. However, in most cases the two curves are nearly indistinguishable over the entire range in radius of the star. 

\hfill

\bibliographystyle{aasjournal}
\bibliography{nixon}

\end{document}